\documentclass[lettersize,journal]{IEEEtran}
\usepackage{amsmath,amsfonts}
\usepackage{algorithmic}
\usepackage{algorithm}
\usepackage{array}
\usepackage[caption=false,font=normalsize,labelfont=sf,textfont=sf]{subfig}
\usepackage{textcomp}
\usepackage{stfloats}
\usepackage{url}
\usepackage{amsmath}
\usepackage{verbatim}
\usepackage{graphicx}
\usepackage{hyperref}
\usepackage{multirow}
\usepackage[table]{xcolor}

\usepackage{cleveref}
\usepackage{booktabs}
\usepackage{cite}
\DeclareMathOperator*{\argmin}{arg\,min}
\usepackage{amssymb}% http://ctan.org/pkg/amssymb
\usepackage{pifont}% http://ctan.org/pkg/pifont
\newcommand{\xmark}{\ding{55}}%

\usepackage{amsmath,amsfonts,bm}

\def\eqref#1{equation~\ref{#1}}
\def\1{\bm{1}}

\def\vc{{\bm{c}}}

\def\ve{{\bm{e}}}

\def\vv{{\bm{v}}}

\def\vx{{\bm{x}}}
\def\vy{{\bm{y}}}

\def\mE{{\bm{E}}}

\def\mX{{\bm{X}}}
\def\mY{{\bm{Y}}}

\DeclareMathAlphabet{\mathsfit}{\encodingdefault}{\sfdefault}{m}{sl}
\SetMathAlphabet{\mathsfit}{bold}{\encodingdefault}{\sfdefault}{bx}{n}

\begin{document}

\title{SemantiCodec: An Ultra Low Bitrate Semantic Audio Codec for General Sound}

% waveform codec VS semantic codec.

\author{Haohe Liu, Xuenan Xu, Yi Yuan, Mengyue Wu, Wenwu Wang, Mark D. Plumbley\\ 
% \url{https://audioldm.github.io/audioldm2/}
\thanks{Haohe Liu, Yi Yuan, Wenwu Wang, and Mark D. Plumbley are with the Centre for Vision, Speech and Signal Processing (CVSSP), University of Surrey, Guilford, UK. Email: \{haohe.liu, yi.yuan, w.wang, m.plumbley\}@surrey.ac.uk.}
\thanks{Xuenan Xu, and Mengyue Wu are with the Department of Computer Science and Engineering, Shanghai Jiao Tong University, Shanghai, China. Email: {\{wsntxxn, mengyuewu\}}@sjtu.edu.cn}
% \thanks{Qiuqiang Kong is with the Department of Electronic Engineering, Chinese University of Hong Kong, Hong Kong, China. Email: {qqkong@ee.cuhk.edu.hk}}
}



\maketitle

\begin{abstract}

Large language models (LLMs) have significantly advanced audio processing through audio codecs that convert audio into discrete tokens, enabling the application of language modelling techniques to audio data. However, traditional codecs often operate at high bitrates or within narrow domains such as speech and lack the semantic clues required for efficient language modelling. Addressing these challenges, we introduce SemantiCodec, a novel codec designed to compress audio into fewer than a hundred tokens per second across diverse audio types, including speech, general sound, and music, without compromising quality. 
\textcolor{black}{SemantiCodec features a dual-encoder architecture: a semantic encoder using a \textcolor{black}{self-supervised} \textcolor{black}{pre-trained Audio Masked Autoencoder~(AudioMAE)}, discretized using k-means clustering on extensive audio data, and an acoustic encoder to capture the remaining details. The semantic and acoustic encoder outputs are used to reconstruct audio via a diffusion-model-based decoder.}
SemantiCodec is presented in three variants with token rates of $25$, $50$, and $100$ per second, supporting a range of ultra-low bit rates between $0.31$~kbps and $1.40$~kbps.  
Experimental results demonstrate that SemantiCodec significantly outperforms the state-of-the-art Descript codec on reconstruction quality. Our results also suggest that SemantiCodec contains significantly richer semantic information \textcolor{black}{than all evaluated state-of-the-art audio codecs}, even at significantly lower bitrates.
Our code and demos are available at \url{https://haoheliu.github.io/SemantiCodec/}.

\end{abstract}

\begin{IEEEkeywords}
audio codec, semantic, low bitrate
\end{IEEEkeywords}

% semantic codec's advantage:
% 1. better reconstruction ability under the same bit rate
% 2. better semantic encoding ability whatever bit rates the baseline use
% 3. **much** better semantic encoding ability when only using the first layer, compared with the baseline
% 4. shorter sequence length under the same bit rate
% Due to 3 and 4, semantic codec is better for audio token language modelling

\section{Introduction}

\IEEEPARstart{A}{udio} codecs are used for encoding and decoding digital audio for efficient telecommunications and broadcasting~\cite{pohlmann2000principles}. Traditional audio encoders compress audio by discarding inaudible details to reduce storage and transmission demands~\cite{pohlmann2000principles}. The degree of compression is typically assessed by the bitrate, indicating the amount of data, in bits per second, used to represent the audio signal, with commonly used bitrate ranges such as $128$~kbps to $320$~kbps for MP3~\cite{sterne2012mp3} and $6$~kbps to $510$~kbps \textcolor{black}{for} the Opus codec~\cite{OpusCodec}. 

With the introduction of deep learning, audio codecs have significantly evolved with better audio quality and bitrate efficiency~\cite{zeghidour2021soundstream}. These cutting-edge codecs utilize vector quantization\textcolor{black}{~(VQ)}~\cite{van2017neural} to learn compact codebooks, whose indices are transmitted instead of raw audio data. The sequence of transmitted indices is also referred to as the token sequence. Unlike traditional audio codecs, neural audio codecs typically operate at lower bit rates while maintaining similar audio quality. For instance, Encodec~\cite{defossez2023high} achieves compression at multiple bitrates between $1.5$~kbps and $24$~kbps, the Descript codec~\cite{kumar2024high} operates at $8$~kbps and $16$~kbps, and the HiFi-Codec pushes the boundaries further by reducing the bitrate to $2$~kbps with acceptable quality~\cite{yang2023hifi}. 
% This advancement is increasingly crucial as the surge in online audio content necessitates reduced transmission bandwidth.

% Beyond the role of storing and transmitting audio, audio codecs are also key for audio language modelling~\cite{borsos2023audiolm}. They simplify complex audio waveforms into discrete tokens with significantly shorter lengths, making the training of audio language models as straightforward as training to text-based language models~\cite{touvron2023llama}. By performing next-token prediction on the audio codec sequence with a transformer decoder, AudioLM~\cite{borsos2023audiolm} has shown success on semantically plausible audio continuation. Later some works employ text as additional conditional information for audio language modelling, leading to works such as text-to-audio generation AudioGen~\cite{kreuk2022audiogen}, text-to-music generation with MusicLM~\cite{agostinelli2023musiclm}, MusicGen~\cite{copet2023simple-musicgen} and text-to-speech generation as signified by Valle~\cite{wang2023neural-valle}. Audio understanding has also shown great improvement with the introduction of the audio codec into the language model, as shown by the work LTU~\cite{gong2023listen}. The joint understanding and generation of audio codecs have also been achieved by AudioPaLM, which fuses the text-based and audio-based language models into one system. Recently, the audio codec has also shown progress on speech quality enhancement tasks~\cite{wang2024selm}.

Beyond the fundamental role of storing and transmitting audio, audio codecs have emerged as critical components in the domain of audio language modelling~\cite{borsos2023audiolm}. Similar to the tokenizers used in text processing~\cite{toraman2023impact}, the neural audio codecs simplify complex audio waveforms into discrete integer tokens with substantially shorter lengths and align the training process of audio language models closely with the training methodologies of text-based language models~\cite{touvron2023llama}. This simplification has enabled models like AudioLM~\cite{borsos2023audiolm} to perform next-token prediction on audio codec sequences, demonstrating success in generating semantically plausible audio continuations. Further advancements have explored the incorporation of text as conditional information, which has paved the way for innovative applications such as text-to-audio generation with AudioGen~\cite{kreuk2022audiogen}, text-to-music generation through MusicLM~\cite{agostinelli2023musiclm} and MusicGen~\cite{copet2023simple-musicgen}, and enhanced text-to-speech systems exemplified by VALL-E~\cite{wang2023neural-valle}. The integration of audio codecs into language models has also significantly advanced audio understanding capabilities, as demonstrated by the LTU model~\cite{gong2023listen}. Moreover, developing systems like AudioPaLM~\cite{rubenstein2023audiopalm} demonstrates the potential of the joint understanding and generation of speech, merging text-based and audio-based language models into a unified framework. 
% Recent innovations have further extended the utility of audio codecs into speech quality enhancement tasks~\cite{wang2024selm}.

\begin{figure}[tbp]
    \centering
    \includegraphics[width=1.0\linewidth]{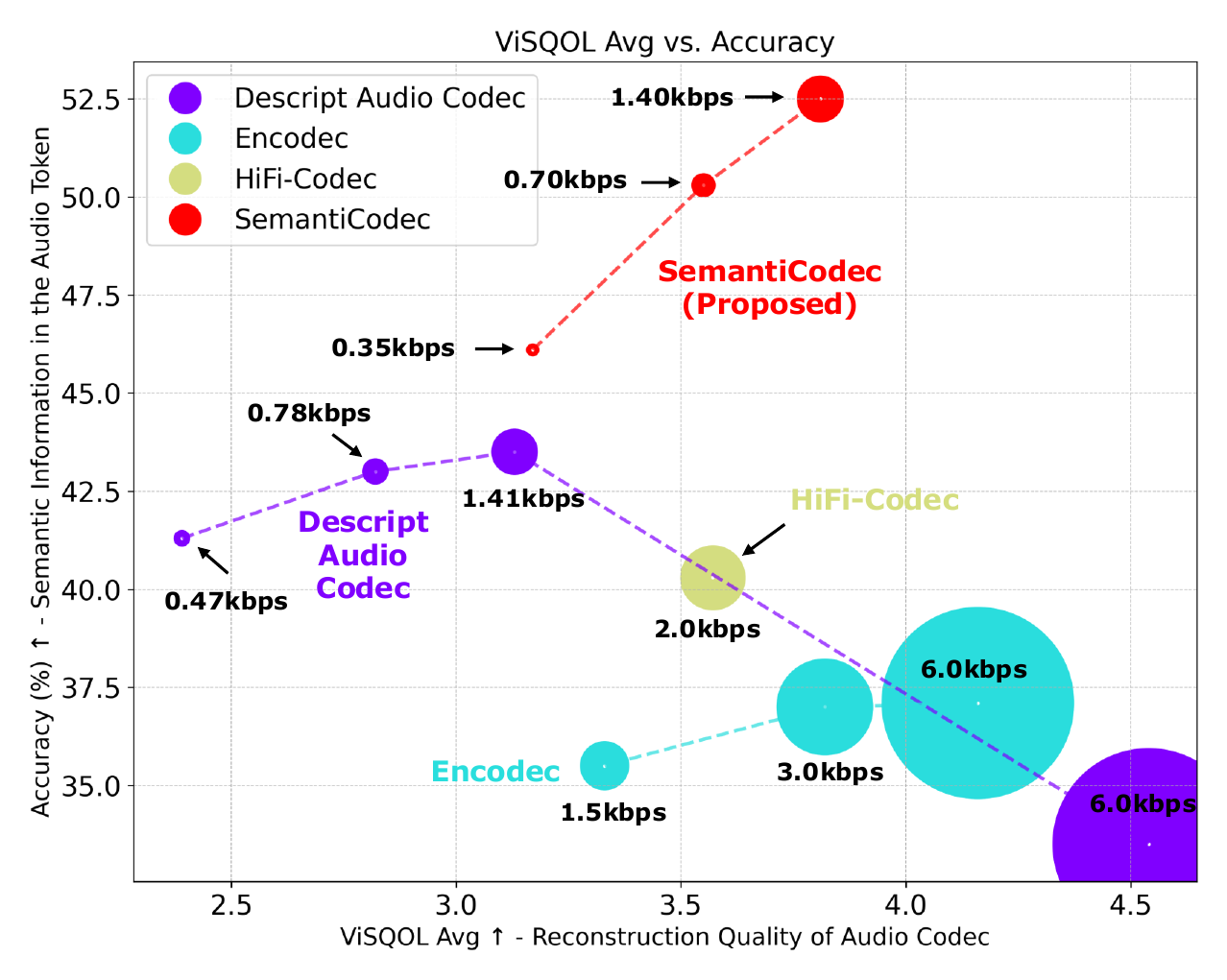}
    \caption{Comparison between SemantiCodec and state-of-the-art codecs. \textcolor{black}{Higher} values on the horizontal and vertical axis indicate better reconstruction quality and semantic information, respectively. The size of circles indicates bitrates, where smaller ones denote lower bitrates.}
    \label{fig:main-semantic}
\end{figure}

Despite the advancement in audio language modelling, the token rate of audio codecs has become a growing concern. For example, the token rate for a $6$~kbps Descript codec is $600$ per second \textcolor{black}{(using six codebooks of size $1024$)}.
The auto-regressive (AR) nature of token generation means that the inference time of an audio language model scales with the \textcolor{black}{rate} of codec tokens, posing challenges not only in computational efficiency but also in model training, where longer sequences demand more computational resources. 
Additionally, longer sequences may lead to challenges on long-term dependencies.
Studies on long-context language models~\cite{sun2021long} indicate that language models do not effectively leverage long-term context, often superficially utilising it.
While low-bitrate audio codecs are available, \textcolor{black}{like the $1.5$ kbps version of Encodec}~\cite{defossez2023high}, their reconstruction quality, with strong artefacts introduced, often falls short of production standards.
This situation underscores a crucial trade-off between efficiency and quality in audio codec development, highlighting the need for codecs that can achieve high-quality audio reconstruction at low bitrates.

The presence and richness of semantic information within token sequences play a crucial role in the learning of language models~\cite{brychcin2014semantic, bayer2016semantic}. 
Support for this assertion also comes from findings by Toraman et al.~\cite{toraman2023impact}, who demonstrated that on six text processing tasks tokenizers operating at a higher level of granularity, such as byte pair encoding~\cite{shibata1999byte}, could substantially outperform character-level tokenizers, which often require the model to expand capacity for understanding.
% to achieve higher-level representations.
Despite these insights on the importance of semantic information in the tokens, our initial investigations reveal that existing audio codecs, already employed in audio language modelling, fall short of capturing adequate semantic information, even at relatively high bitrate settings.
For instance, when employing the latent encodings of the $6.0$~kbps Descript codec~\cite{kumar2024high} for classification tasks across seven benchmarks in the HEAR benchmark~\cite{turian2022hear}, the average accuracy is only $33\%$, as demonstrated in \Cref{subsec:semantic_result}.
In contrast, without fine-tuning, a self-supervised pretrained AudioMAE encoder~\cite{huang2022masked} results in a significantly higher accuracy of $61\%$.
The classification accuracy is the indicator of semantic richness within codecs, which is crucial for audio language modelling.
Moreover, our analysis indicates that using just the first vector quantization layer of the $6.0$~kbps Descript codec, which is often considered to be the most important layer and selected for audio language modelling~\cite{wang2023neural-valle}, the classification accuracy drops even further to an average of $25\%$.
This deficiency in capturing and encoding rich semantic information in audio codecs can potentially hinder the performance of audio language models.
% , underscoring a pressing need for advancements in codec design that prioritize semantic depth and clarity. 
% TODO say this: We show more details of audio codec semantic information evaluation in Section~\ref{none}.

% This paper proposes SemantiCodec, which addresses the challenges of long sequence length and the lack of semantic information by utilizing the strong generative modelling ability of the diffusion model and the representation learnt by the self-supervised pretrained AudioMAE. 
In this paper, we introduce SemantiCodec, a novel audio codec designed to tackle the issues of excessive token rate and insufficient semantic encoding in current audio codecs.
SemantiCodec exhibits richer semantic information and similar reconstruction quality with lower bitrates than previous codecs, as illustrated in \Cref{fig:main-semantic}. 
Our approach uses the strong generative capabilities of diffusion models alongside the rich audio representations learned by the self-supervised AudioMAE. 
SemantiCodec processes mel-spectrograms through two encoders sequentially with two distinct vector quantization~(VQ) layers. The first VQ layer is constructed using centroids derived from k-means clustering~\cite{macqueen1967some} performed on a large dataset of AudioMAE features, ensuring the capture of semantic information. In contrast, the second layer employs a conventional learnable VQ mechanism, enhancing the audio reconstruction fidelity of SemantiCodec.
% quantized AudioMAE feature to reconstruct audio with high fidelity.
Our empirical analysis reveals that the semantic content is predominantly encoded within the first VQ layer, 
% accounting for over $95\%$ of the semantic information. 
contributing over $95\%$ to the classification accuracy.
However, integrating the second VQ layer is critical in significantly elevating the quality and intelligibility of the reconstructed audio. 
The concatenated quantized outputs from both VQ layers serve as the conditional input for a latent diffusion model~(LDM), which follows the architecture of AudioLDM~\cite{liu2023audioldm}. The latent diffusion model utilizes the provided conditions to reconstruct high-quality audio after quantization. \textcolor{black}{Our experiment shows that SemantiCodec, at similar bitrates, outperforms the state-of-the-art audio codecs on reconstruction quality and contains significantly richer semantic information in the} audio tokens, leading to improved classification accuracy in audio understanding. In summary, our contributions are listed as follows:

\begin{itemize}
    \item We propose SemantiCodec, which leverages strong generative models and the rich features learnt by \textcolor{black}{a} self-supervised model for semantic-driven audio encoding and reconstruction.
    \item \textcolor{black}{To our knowledge, \textcolor{black}{SemantiCodec is the first work to use a diffusion based decoder in a neural audio codec at the time of writing}.}
    \item SemantiCodec achieves strong reconstruction performance across general sound types at exceptionally low token rates of $25$, $50$, and $100$ per second, surpassing counterparts operating at significantly higher token rates.
    \item Evaluation on audio classification benchmarks demonstrates the significantly richer semantic information in the token\textcolor{black}{s} of SemantiCodec, even with a single layer of vector quantization, indicating strong potential in future audio language modelling with the SemantiCodec.
    % \item Comparative studies show that SemantiCodec outperforms leading neural audio codecs in reconstruction quality at comparable bitrates.
\end{itemize}

\section{Related Work}

\subsection{Neural Audio Codecs}

\noindent
Traditional audio codecs, as detailed by Valin et al.~\cite{valin2012definition} and Dietz et al.~\cite{dietz2015overview}, have demonstrated their ability to achieve low latency audio compression across a variety of audio types. However, these traditional approaches often fail to deliver high-quality audio reconstruction at low-bitrate settings, such as $3.0$ kbps.
% The advent of deep learning has spurred the development of a new wave of audio codecs based on neural network technologies, offering significantly improved fidelity in audio reconstruction at comparable bitrates.
Pioneering deep-learning-based audio codecs, Garbacea et al.~\cite{garbacea2019low} introduce the use of vector-quantized variational auto-encoders~(VQ-VAEs) for learning neural codecs tailored to speech data. SoundStream~\cite{zeghidour2021soundstream} proposes a universal codec adaptable to various audio types, incorporating a residual vector quantization~(RVQ) module to enhance the quantization process and a generative adversarial network~(GAN)~\cite{creswell2018generative} to improve the reconstruction quality. 
% setting a precedent for the encoder-RVQ-decoder architecture and VQ-GAN training methodology that would be adopted by subsequent codecs.
Following a similar trajectory, Encodec~\cite{defossez2023high} advances the capabilities of SoundStream by integrating multi-scale discriminators and a loss-balancing strategy for reconstruction alongside an additional language model to facilitate further compression. 
HiFi-Codec~\cite{yang2023hifi} introduced group-residual vector quantization~(GRVQ), a novel approach aimed at reducing the number of codebooks required while preserving the reconstruction quality. Descript codec~\cite{kumar2024high} offers enhancements to Encodec, achieving significantly \textcolor{black}{better} reconstruction performance. 
% \textcolor{black}{A few months after the writing of this paper, Music2Latent~\cite{pasini2024music2latent} was introduced, utilizing a consistency model~\cite{pmlr-v202-song23a} for audio compression to achieve high-fidelity, single-step audio reconstruction.}

To compare our proposed SemantiCodec with previous systems, our experimental analysis primarily focuses on evaluating the richness of semantic information and the quality of audio reconstruction.

\subsection{Semantic Audio Representation Learning}

\label{sec: semantic_audio_representation-learning}
Self-supervised learning~(SSL) has exhibited strong performance in audio representation learning.
SSL models can be categorized into two types based on their pre-training tasks: discriminative SSL and reconstructive SSL.
Discriminative SSL models, exemplified by HuBERT~\cite{hsu2021hubert}, COLA~\cite{saeed2021contrastive}, and BEATs~\cite{chen2023beats}, employ strategies such as contrastive learning to differentiate between positive and negative pairs, or masked language modelling (MLM) techniques to predict the quantized labels of masked segments, leveraging contextual information.
Conversely, reconstructive SSL models, such as AudioMAE~\cite{huang2022masked}, take inspiration from MLM principles but pivot towards reconstructing the original audio content from masked segments. 
% These features, compared with raw waveforms or spectrograms, serve as more efficient semantic representations, indicated by their stronger performance on downstream understanding tasks.
Given the reconstructive nature of AudioMAE pre-training, the AudioMAE features are potentially more balanced in acoustic and semantic information than those derived from only discriminative pre-training. 
In this work, we develop SemantiCodec with an AudioMAE encoder as one of the fundamental components.

\begin{figure*}[t!]
    \centering
    \includegraphics[width=0.9\textwidth]{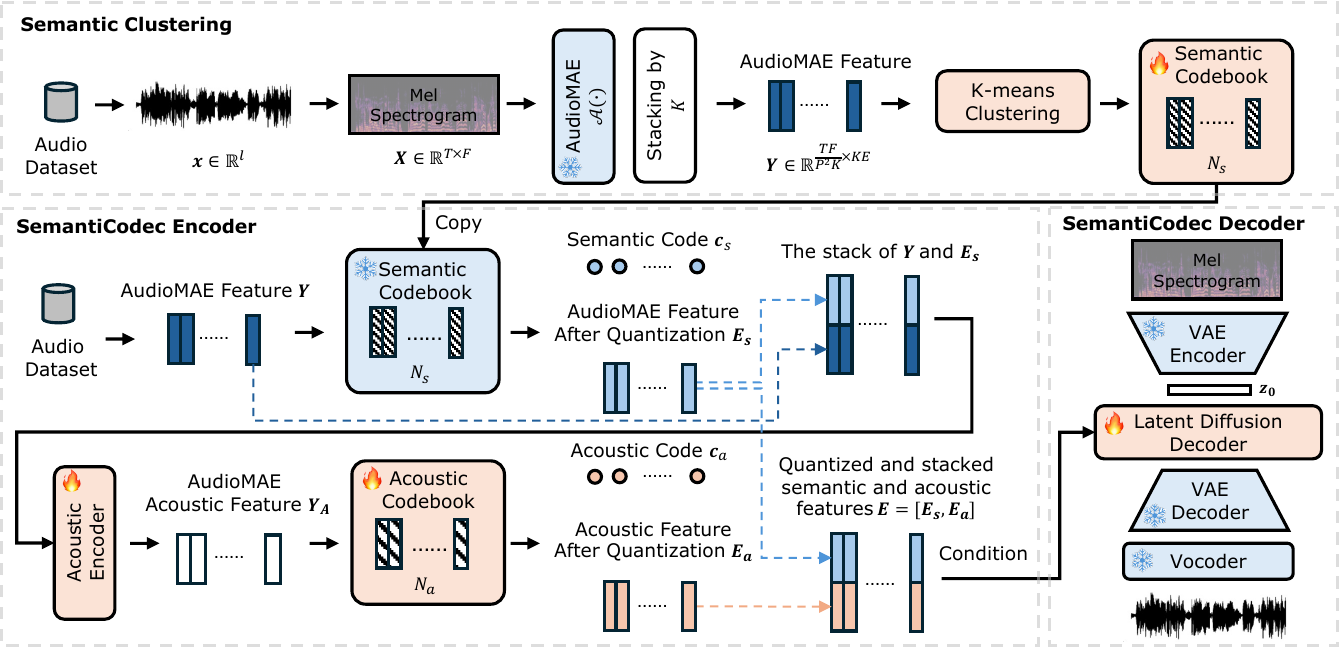}
    \caption{
    SemantiCodec architecture. For an input audio clip, quantized semantic representation $\mE_{s}$ is obtained via a \textcolor{black}{pre-computed codebook} using k-means clustering on the AudioMAE embeddings. Then $\mY$ and $\mE_\text{s}$ are concatenated and fed to a residual encoder to complement acoustic details, which is discretized to $\mE_{a}$ by a vector quantization module. SemantiCodec encoder output $\mE$ is obtained by concatenating $\mE_{s}$ and $\mE_{a}$. A latent diffusion model is trained to generate the original audio clip conditioned on $\mE$. The snowflake and fire symbols denote frozen and learnable parameters, respectively.
    }
    \label{fig:model_architecture}
\end{figure*}

\subsection{Conditional Audio Generation}
% diffusion
% With the development of generative models~\cite{goodfellow2014generative,kingma2013auto,rezende2015variational}, conditional audio generation has become increasingly popular for different types of audio.
% Speech generation~\cite{popov2021gradtts,kim2021conditional} is conditioned on transcriptions and styles (e.g., speaker, emotion, prosody) while music and sound effect generation is conditioned on textual descriptions~\cite{kreuk2022audiogen,liu2023audioldm} or visual signals~\cite{iashin2021taming-specvqgan,sheffer2023hear}.
% Among these studies, diffusion models~\cite{ho2020denoising} have demonstrated strong generative modelling performance.
% They can generate diverse data samples with a high quality and are easier to train than GANs.
% To reduce the computational complexity in training and inference, latent diffusion models (LDMs)~\cite{rombach2022high-stablediffusion} are proposed to train diffusion models on low-dimensional latent distribution, which is obtained from a variational auto-encoder (VAE).
% LDMs are successful in many conditional audio generation works~\cite{liu2023audioldm,ghosal2023text-tango,huang2023make-an-audio}.
% Specifically, AudioLDM 2~\cite{liu2023audioldm2} is trained to generate audio samples conditioned on AudioMAE embeddings.
% Therefore, in this work, we adopt an LDM as the reconstruction model to generate the original audio sample conditioned on our proposed semantic-rich tokens.

The development of generative models~\cite{goodfellow2014generative, kingma2013auto, rezende2015variational} has significantly propelled the field of conditional audio generation. Speech generation technologies~\cite{popov2021gradtts,kim2021conditional,tan2022naturalspeech} have evolved to produce speech conditioned on transcriptions and specific styles, such as speaker identity, emotion, and prosody. The generative model has also enabled novel tasks such as binaural sound synthesis~\cite{leng2022binauralgrad} and synthetic speech quality refinement~\cite{ResGrad}. Meanwhile, the generation of music and sound effects has been extended to be conditioned on textual descriptions~\cite{kreuk2022audiogen,liu2023audioldm, chen2024musicldm} and visual cues~\cite{iashin2021taming-specvqgan,sheffer2023hear}, demonstrating the versatility of generative architectures.

Diffusion models~\cite{ho2020denoising} stand out for their exceptional generative capabilities in producing diverse and high-quality samples. Diffusion models offer a more tractable training process than GANs, emerging as a preferred choice for many researchers. To address the computational demands associated with training and inference in high-dimensional spaces, latent diffusion models\textcolor{black}{~(LDM)}~\cite{rombach2022high-stablediffusion} have been introduced. LDMs operate on a lower-dimensional latent space derived from a VAE, significantly reducing computational complexity while maintaining the generative power of traditional diffusion models. This approach has been successfully applied in various conditional audio generation models, including AudioLDM~\cite{liu2023audioldm}, TANGO~\cite{ghosal2023text-tango}, AudioSR~\cite{liu2024audiosr} and Make-An-Audio~\cite{huang2023make-an-audio}, illustrating the flexibility of the diffusion model. \textcolor{black}{In this work, we leverage the strong generative modelling ability of LDMs to build a SemantiCodec decoder.}

% A notable implementation, AudioLDM 2~\cite{liu2023audioldm2}, leverages AudioMAE embeddings as a condition for generating audio samples, showcasing the potential of combining advanced encoding techniques with diffusion-based generation. Inspired by these advancements, our work adopts an LDM for the reconstruction phase, generating original audio samples conditioned on semantically rich tokens proposed by our SemantiCodec. This strategic integration aims to harness the strengths of LDMs in producing high-quality, diverse audio content, thus enhancing the overall performance and capability of our codec.

\section{SemantiCodec}

\subsection{System Overview}
\label{sec: overview}
As shown in Figure~\ref{fig:model_architecture}, given an input audio \(\vx \in \mathbb{R}^l\), where \(l\) denotes the sample length of audio, we initially transform \(\vx\) into the mel-spectrogram \(\mX \in \mathbb{R}^{T \times F}\), with \(T\) and \(F\) indicating the temporal and frequency dimensions, respectively. Leveraging a pretrained AudioMAE \(\mathcal{A}(\cdot)\), we compute the AudioMAE features \(\tilde{\mY} = \mathcal{A}(X) = [\tilde{\vy}_1, \tilde{\vy}_2, \ldots, \tilde{\vy}_L] \in \mathbb{R}^{L \times E}\), where \(L = \frac{TF}{P^2}\) denotes the number of patch embedding vectors, and \(P\) and \(E\) represent the patch size and embedding size of AudioMAE, respectively. Each patch is a distinct, non-overlapping block of the mel-spectrogram processed by the AudioMAE, with multiple patches collectively forming the input to the AudioMAE encoder.

To reduce the number of patch embedding vectors, which directly influence the bitrate after quantization, we aggregate adjacent vectors of \(\tilde{\mY}\) into \(\mY = [\vy_1, \vy_2, \ldots, \vy_{\frac{L}{K}}] \in \mathbb{R}^{\frac{L}{K} \times KE}\), where \(K \in \{1, 2, 4\}\) is the stack factor, yielding \textcolor{black}{\(\vy_i = [\tilde{\vy}_{iK}, \ldots, \tilde{\vy}_{(i+1)K-1}]\) for \(i \in \{0, 1, \ldots, \frac{L}{K}-1\}\). }
Following extensive clustering on the vector \(\vy_i\), we derive the semantic codebook \(\mathbb{E}_\text{s} = [\ve_1, \ve_2, \ldots, \ve_{N_\text{s}}]\), where $N_\text{s}$ denotes the number of entries in the semantic codebook. We refer to this clustering process as \textit{semantic clustering}.

The stacked feature \(\mY\) undergoes initial quantization by \(\mathbb{E}_\text{s}\) into semantic tokens \(\vc_\text{s}\) and semantic feature \(\mE_\text{s} \in \mathbb{R}^{\frac{L}{K} \times KE} \).
Subsequently, we concatenate \(\mY\) and \(\mE_\text{s}\), employing an acoustic encoder \(\mathcal{F}(\cdot)\) to compute the acoustic feature \(\mY_\text{A}\), which is then quantized via an acoustic vector quantization layer with entries \textcolor{black}{\(\mathbb{E}_\text{a} \in \mathbb{R}^{N_\text{a} \times KE}\)}, outputting the acoustic tokens \(\vc_\text{a}\) and quantized acoustic feature \(\mE_\text{a}\). The final tokens for the input audio $\vx$ is a merge of semantic and acoustic tokens: \(\vc = [\vc_\text{s}, \vc_\text{a}]\). 

The decoder of SemantiCodec utilizes a latent diffusion model conditioned on the concatenated quantized semantic and acoustic features \(\mE = [\mE_\text{s}, \mE_\text{a}]\). The estimation of the latent diffusion model is further decoded back to waveform by a pretrained VAE decoder and a mel-spectrogram vocoder. The acoustic encoder \(\mathcal{F}(\cdot)\) is joint-optimized with the acoustic codebook $\mathbb{E}_\text{a}$ and the latent diffusion model.

% Based on the AudioMAE feature, the SemantiCodec encoder performs two layers of vector quantization~(VQ), in which the semantic codebook from the first VQ layer is pre-computed using k-means clustering and the acoustic codebook in the second VQ layer is learnable. 

\subsection{Semantic Clustering}

% AudioMAE features have shown potential for preserving both rich semantic and acoustic information. As explored in Section~\ref{sec: semantic_audio_representation-learning}, the AudioMAE feature is demonstrated to be well-suited for reconstruction tasks compared with other discrimination loss based self-supervised pretrained models. Previous studies~\cite{liu2023audioldm2} have also confirmed that the AudioMAE feature encapsulates sufficient detail for accurate audio reconstruction. Besides, the AudioMAE feature has also demonstrated strong performance on the downstream classification benchmarks~\cite{huang2022masked}. Therefore, we choose to utilize the AudioMAE feature as the input for the SemantiCodec encoder to facilitate both audio reconstruction and semantic information preservation.

AudioMAE features stand out for their ability to preserve semantic and acoustic information~\cite{liu2023audioldm2}, positioning them as highly effective features for reconstruction tasks. 
% As detailed in Section~\ref{sec: semantic_audio_representation-learning}, these features are potentially superior for reconstruction purposes compared to contrastive loss features. 
% This assertion is supported by previous research~\cite{liu2023audioldm2}, which verifies the capability of AudioMAE features to enable precise audio reconstruction. 
Additionally, AudioMAE features have demonstrated strong performance in downstream classification benchmarks~\cite{huang2022masked}. Given these attributes, the AudioMAE feature is selected as the input for the SemantiCodec encoder, aiming to optimize audio reconstruction quality while ensuring the retention of semantic content.

We follow AudioLDM 2~\cite{liu2023audioldm2} for AudioMAE feature extraction. Given an audio mel spectrogram representation $\mX \in \mathbb{R}^{T \times F}$, the AudioMAE first transforms $\mX$ into patches of dimensions $P \times P$. These patches form the inputs to the AudioMAE encoder, which leverages a design akin to the vision transformer~\cite{dosovitskiy2020image}.
The output $\mY_0$ of the AudioMAE encoder has a dimension of $\frac{T}{P} \times \frac{F}{P} \times E$, which can be viewed as a sequence of tensors with length $L=\frac{TF}{P^2}$ and an embedding dimension of $E$. We stack the adjacent $K$ frames of $\mY_0$ on the embedding dimension to form stacked AudioMAE feature vectors \(\mY = \{\vy_1, \vy_2, \ldots, \vy_{\frac{L}{K}}\}\), which are used both for semantic clustering and as the input of SemantiCodec encoder.

To perform semantic quantization on the AudioMAE feature vectors \(\vy_i\), we utilize k-means clustering~\cite{macqueen1967some}, a widely used technique for partitioning a dataset into clusters in which each data point belongs to the cluster with the nearest mean. Our preliminary experiment indicates that the diverse acoustic characteristics of different audio types could lead to suboptimal outcomes when a single k-means clustering model is applied to a varied audio dataset. For instance, speech may necessitate finer granularity in clustering due to its rich semantic content, unlike more homogeneous sounds such as wind or church bells. We employ an ensemble clustering approach to overcome these issues and enhance semantic clustering accuracy, motivated by the cluster ensemble approach used in HuBERT~\cite{hsu2021hubert}. This involves training distinct k-means models for three specific audio categories: speech, music, and general sounds. The codebooks \textcolor{black}{(i.e., k-means cluster centroids)} derived from these domains are then \textcolor{black}{combined together}, forming a \textcolor{black}{new} ensembled codebook that accommodates the distinct acoustic features of each audio category.

% The objective of k-means clustering is to categorize the large-scale dataset of AudioMAE feature vectors into \(N_\text{s}\) clusters, each represented by a centroid \(e_j\) that corresponds to the mean of the vectors in the cluster.

% Given a set of feature vectors \(\mY = \{\vy_1, \vy_2, \ldots, \vy_{\frac{L}{K}}\}\), the k-means clustering algorithm seeks to minimize the within-cluster sum of squares (WCSS), given by

% where \(\mathbb{E}_\text{s} = \{\ve_1, \ve_2, \ldots, \ve_{N_\text{s}}\}\) represents the set of cluster centroids. The objective function is to adjust the positions of the centroids to minimize the sum of the distances of all points to their closest centroid.

% The heterogeneity in the acoustic characteristics across different audio types can lead to sub-optimal clustering performance when a unified k-means clustering model is trained on diverse audio types.
% To circumvent such challenges and enhance the granularity of semantic clustering, we adopt a stratified clustering approach.
% We separately train k-means clustering models on three distinct audio categories, including speech, music, and sound effects.
% Codebooks from the three clustering models are merged to form a unified semantic clustering model.

\subsection{SemantiCodec Encoder}
As introduced in Section~\ref{sec: overview}, the encoding process of the SemantiCodec includes taking the stacked AudioMAE feature $\mY$ as input and calculating the tokens $\vc=[\vc_\text{s},\vc_\text{a}]$ and the latent features $\mE=[\mE_\text{s}, \mE_\text{a}]$.
This part introduces the derivation of $\vc$ and $\mE$.

\subsubsection{Semantic Encoder}

With the $N_\text{s}$ semantic codebook centroids $\ve_j \in \mathbb{E}_\text{s} = \{\ve_1, \ve_2, \ldots, \ve_{N_\text{s}}\}$, the semantic quantization process is given by

\begin{equation}
\label{eq:kmeans}
% c_\text{s}(i), \mE_\text{s}(i) = \argmin_{\ve_j \in \mathbb{E}_\text{s}} \| \vy_i - \ve_j \|^2, \quad i \in \left\{1, 2, \ldots, \frac{L}{K}\right\},
% \small
c_\text{s}(i) = \argmin_{j \in \{1, \ldots, N_\text{s}\}} \| \vy_i - \ve_j \|^2, \quad \mE_\text{s}(i) = \ve_{c_\text{s}(i)},
\end{equation}

\noindent
where \textcolor{black}{\(i \in \{0, 1, \ldots, \frac{L}{K}-1\}\)}, \(c_\text{s}(i)\) denotes the index of the closest centroid in the semantic codebook \(\mathbb{E}_\text{s}\) to the feature vector \(\vy_i\), and \(\mE_\text{s}(i)\) is the centroid vector. This quantization step effectively maps each high-dimensional feature vector \(\vy_i\) into a discrete semantic token \(c_\text{s}(i)\) and corresponding quantized semantic feature \(\mE_\text{s}(i)\).

\subsubsection{Acoustic Encoder}
While the quantized AudioMAE feature $\mE_\text{s}$ encapsulates rich semantic information, our experiments indicate that using $\mE_\text{s}$ alone as the condition for audio reconstruction leads to sub-optimal quality~(see \Cref{tab:reconstruction_result}), such as unintelligible speech.
To address this, we introduce an additional acoustic encoder module with vector quantization to capture discrete acoustic detail-oriented representations.
The input feature we used for acoustic quantization is calculated by an acoustic encoder $\mathcal{F}_{\Phi}(\cdot)$, which takes the stack of the AudioMAE feature before and after semantic quantization as input and outputs the acoustic feature, given by

\begin{equation}
    \mY_{A} = \mathcal{F}_{\Phi}([\mY, \mE_\text{s}]) \in \mathbb{R}^{\frac{L}{K}\times EK},
\end{equation}

\noindent
where $\Phi$ denotes the trainable parameter in the acoustic encoder. Specifically, we employ a bi-directional long short-term memory (BiLSTM) based network~\cite{sak2014long} as the acoustic encoder.
\textcolor{black}{We use both $\mY$ and $\mE_\text{s}$ as inputs because $\mE_\text{s}$ undergoes information loss during semantic quantization, while $\mY$ retains most of the necessary information for audio reconstruction. By providing both $\mY$ and $\mE_\text{s}$, the second acoustic quantization layer can effectively compensate for the information lost during semantic quantization.}
% TODO xnx 要不要提 the same embedding dimension
% The output of the acoustic encoder has the same embedding dimension as stacked AudioMAE embedding. 

We use the quantization of \(\mY_{A}\) to convert the detailed acoustic nuances into a compact, discrete format.
We define the acoustic codebook as \(\mathbb{E}_\text{a} = \{\ve_1, \ve_2, \ldots, \ve_{N_\text{a}}\}\), where \(N_\text{a}\) denotes the number of codebook entries and each \(\ve_i \in \mathbb{R}^{EK}\) represents an individual codebook vector.
For each vector \(\vy_{a,i} \in \mY_{A}\), the quantization process is as follows:

\begin{equation}
\label{eq:acoustic_vq}
% c_\text{a}(i), \mE_\text{a}(i) = \argmin_{\ve_j \in \mathbb{E}_\text{a}} \| \vy_{a,i} - \ve_j \|^2, \quad i \in \{1, 2, \ldots, \frac{L}{K}\},
% \small
c_\text{a}(i) = \argmin_{j \in \{1,\ldots, N_\text{a}\}} \| \vy_{a,i} - \ve_j \|^2, \quad \mE_\text{a}(i) = \ve_{c_\text{a}(i)},
\end{equation}

\noindent
where \textcolor{black}{\(i \in \{0, 1, \ldots, \frac{L}{K}-1\}\)}, \(c_\text{a}(i)\) identifies the index of the nearest centroid within the acoustic codebook \(\mathbb{E}_\text{a}\) for the feature vector \(\vy_{a,i}\), and \(\mE_\text{a}(i)\) denotes the quantized vector.
To encourage the acoustic encoder to produce representations that closely match the codebook entries and stabilise training, we employ a commitment loss~\cite{van2017neural}, given by

\begin{equation}
\label{eq:commitment_loss}
\mathcal{L}_\text{commit} = \sum_{i} \| \vy_{a,i} - \mE_\text{a}(i) \|^2.
\end{equation}

Following the success of HiFi-Codec~\cite{yang2023hifi}, our approach utilizes an exponential moving average (EMA) mechanism for codebook update. The final tokens $\vc$ and the representation $\mE$ are the concatenation of output\textcolor{black}{s} from both the semantic quantization layer and acoustic quantization layer, given by

\begin{equation}
    \vc=[\vc_\text{s},\vc_\text{a}]\in \mathbb{N}^{\frac{2L}{K}}, \mE=[\mE_\text{s}, \mE_\text{a}] \in \mathbb{R}^{\frac{L}{K}\times 2EK}.
\end{equation}

Note that the dual-layer vector quantization architecture (semantic quantization and acoustic quantization) in SemantiCodec is different from the residual vector quantization (RVQ) approach~\cite{zeghidour2021soundstream, defossez2023high}, where the second layer \textcolor{black}{aims to refine the residual from the first VQ layer}.
% \textcolor{black}{Although both our method and RVQ rely on multiple layers of vector quantization,} our approach allows the output of the second layer to be optimized in conjunction with the diffusion decoder for end-to-end training, \textcolor{black}{without relying on residual calculations between VQ layers}.
\textcolor{black}{Although both SemantiCodec and RVQ rely on multiple layers of vector quantization, SemantiCodec does not explicitly calculate residuals between each layer. Instead, SemantiCodec concatenates the outputs of each VQ layer to form the output of the encoder.}

\textcolor{black}{Our experiments in Section~\ref{sec: results} demonstrate that the dual-layer vector quantization in SemantiCodec effectively decouple the semantic and acoustic information. Specifically, removing the acoustic quantization layer significantly degrades the reconstruction quality~(Table~\ref{tab:reconstruction_result}), while having only a minor impact on the semantic information in the latent space~(Table~\ref{tab:semantic_result}).}

\subsection{Latent Diffusion Model for Reconstruction}
Following the success of latent diffusion models~(LDMs) for conditional audio generation~\cite{liu2024audiosr, liu2023audioldm, huang2023make-an-audio}, we employ an LDM as the decoder to reconstruct the original audio $x$.
The LDM models the data distribution in a latent space constructed from a VAE, which is directly adopted from the VAE used in AudioLDM 2~\cite{liu2023audioldm2}. 
\textcolor{black}{Note that the VAE encoder is only used during SemantiCodec training, and is discarded during SemantiCodec inference.}
Compared with the original diffusion model~\cite{ddpm}, the high-dimensional spectrogram $\mX$ is compressed to the low-dimensional latent $\mathbf{z}_0$ in the LDM to significantly alleviate the computations.
Then a diffusion model is trained to gradually generate $\mathbf{z}_0$ from Gaussian noise.
% It can be formulated as:
% \begin{equation}
% \mathbf{z}_0 = \mathrm{Encoder}_{\mathrm{VAE}}(S).
% \end{equation}
The forward diffusion process comprises a series of $N$ Markov transition steps that gradually transform $\mathbf{z}_0$ into a Gaussian distribution by noise injection.
The forward step $n - 1$ is defined as:
\begin{equation}
    q(\mathbf{z}_n|\mathbf{z}_{n-1}) = \sqrt{1 - \beta_n}\mathbf{z}_{n-1} + \sqrt{\beta_n}\epsilon_n,
\end{equation}
where $\beta_n$ is the predefined noise schedule.
By compositing these forward steps, we can derive the closed-form distribution of an arbitrary step $n$ given the initial $\mathbf{z}_0$~\cite{ddpm}:
\begin{equation}
    q(\mathbf{z}_n|\mathbf{z}_0) = \sqrt{\bar{\alpha}_n}\mathbf{z}_0 + \sqrt{1 - \bar{\alpha}_n}\epsilon_n,
\end{equation}
where $\alpha_n = 1 - \beta_n, \bar{\alpha}_n = \prod_{n=1}^n\alpha_n$ and $\epsilon \sim \mathcal{N}(0, I)$.
With enough diffusion steps $N$, $q(\mathbf{z}_n)$ will approximate a standard Gaussian distribution $ \mathcal{N}(0, I)$. The LDM is trained to model the reverse probability $p_{\theta}(\mathbf{z}_{n-1}|\mathbf{z}_n, \mE)$ conditioned on the SemantiCodec encoder output $\mE$.
\textcolor{black}{Although $\mE_\text{a}$ may already contain part of the information from $\mE_\text{s}$, we feed both into the diffusion model as conditions to enable both quantization layers to collaboratively capture as much information from $\mY$ as possible.}

Recent research by Lin et al.~\cite{lin2024common} has identified limitations in the prevalent noise scheduling techniques employed in diffusion models, specifically highlighting that the noisy latent at the last forward diffusion step $\mathbf{z}_N$ fails to follow a Gaussian distribution.
To rectify this discrepancy, we adopt the strategy proposed by Lin et al.\cite{lin2024common} and implement a cosine noise schedule. This modification guarantees the attainment of a standard Gaussian distribution in the last step of the diffusion process during training, thereby enhancing the consistency of the LDM between training and inference.
% With the rescaling, the distribution at the final diffusion forward step is forced to be a standard Gaussian.
% Therefore, the distribution mismatch at the first step of sampling is eliminated.
% Second, we switch the standard noise prediction to velocity prediction proposed in \cite{salimans2021progressive}.
To improve the generation performance and stabilize the sampling process, we switch the standard noise prediction objective to velocity prediction proposed in~\cite{salimans2021progressive}. The LDM training loss~\cite{salimans2021progressive} can be formulated as:
\begin{align}
    \vv_n &= \sqrt{\bar{\alpha}_n}\epsilon - \sqrt{1 - \bar{\alpha}_n}\mathbf{z}_0, \\
    \mathcal{L}_\text{recon} &= \lVert \vv_n - \mathcal{G}_{\theta}(\mathbf{z}_n, n, \mE)\rVert^2,
    \label{eq:loss}
\end{align}
where $\mathcal{G}_{\theta}$ denotes the LDM and $\theta$ is the set of trainable parameters.

For model inference, we adopt the \textcolor{black}{Denoising Diffusion Implicit Models~(DDIM)} sampler~\cite{song2020denoising-ddim}\textcolor{black}{, which uses non-Markovian diffusion processes for sampling, allowing for computationally efficient sampling while maintaining high generation quality.}
Finally, the audio $\hat{\vx}$ is reconstructed via the pretrained VAE decoder and a vocoder.
The vocoder is a HiFi-GAN-based architecture~\cite{kong2020hifi} and is directly adopted from the pretrained AudioLDM 2.
Our LDM adopts the Transformer-UNet~(T-UNet) introduced in AudioLDM 2~\cite{liu2023audioldm2}, which incorporates self-attention and cross-attention Transformer blocks between convolutional blocks, significantly enhancing the model capacity \textcolor{black}{for} complex audio patterns.
However, our experiments suggest that the impact of T-UNet parameter numbers, such as $75$ million or $346$ million, on the quality of audio reconstruction is relatively minor compared to their role in text-to-audio generation tasks. Therefore, we adopt a T-UNet with reduced parameter size compared with AudioLDM 2.

% Using the velocity-prediction formulation, each step of DDIM sampling is given by:
% \begin{align}
%     \epsilon &= \sqrt{\bar{\alpha}_n}v_{\theta}(\mathbf{z}_n, n, Y^{\prime}_{sem}) + \sqrt{1 - \bar{\alpha}_n}\mathbf{z}_n\\
%     \tilde{z}_0 &= \sqrt{\bar{\alpha}_n}\mathbf{z}_n - \sqrt{1 - \bar{\alpha}_n}v_{\theta}(\mathbf{z}_n, n, Y^{\prime}_{sem})\\
%     \hat{z}_{n-1} &= \sqrt{\bar{\alpha}_{n-1}}\tilde{z}_0 + \sqrt{1 - \bar{\alpha}_{n-1} - \sigma_n^2}\epsilon
% \end{align}

% \subsubsection{Classifier-Free Guidance}

To achieve potentially better quality in reconstruction, we adopt classifier-free guidance~(CFG)~\cite{CFG,Glide}, a common approach to guiding the audio generation \textcolor{black}{in} diffusion models.
The condition $\mE$ in Equation~(\ref{eq:loss}) is randomly discarded with a certain probability during training so that both conditional generation models $v_{\theta}(\mathbf{z}_n, n, \mE)$ and unconditional generation models $v_{\theta}(\mathbf{z}_n, n)$ are optimized in a multi-task paradigm.
During sampling, the original $v_{\theta}(\mathbf{z}_n, n, \mE)$ is complemented by the weighted combination of velocities predicted by conditional and unconditional models:
\begin{equation}
    \label{eq:cfg}
    % \tilde{v}_{\theta}(\mathbf{z}_n, n, \mE) = 
    (1 - w)\cdot v_{\theta}(\mathbf{z}_n, n, \mE) + w\cdot v_{\theta}(\mathbf{z}_n, n),
\end{equation}

\noindent
where $w$ is the guidance scale. We show the effect of different CFG guidance scales in \Cref{fig:ablation}.

\subsection{Training Objective}

As shown in \Cref{fig:model_architecture}, we keep the pretrained AudioMAE, VAE and vocoder parameters frozen.
The k-means semantic clustering centroids are separately obtained before the training of the latent diffusion model and are also kept frozen.
The acoustic encoder, acoustic vector quantization layer and the latent diffusion model are jointly optimized by a sum of the reconstruction loss and the commitment loss in the acoustic vector quantization layer, denoted by

\begin{equation}
    \mathcal{L} = \mathcal{L}_\text{recon} + \mathcal{L}_\text{commit}.
\end{equation}

% Since the quantization operation in \Cref{eq:vector_quantize} is non-differentiable, the straight-through estimator is used to approximate the gradient:
% \begin{align}
%     \begin{split}
%         \mathcal{L}_{commit} &=  \lVert Y_\text{A} - \mathrm{sg}[Y_\text{A}^{\prime}]\rVert^2,\\
%         \mathcal{L}_{code} &= \lVert \mathrm{sg}[Y_\text{A}] - Y_\text{A}^{\prime}\rVert^2,
%     \end{split}
% \end{align}
% where $\mathrm{sg}[\cdot]$ denotes the stop-gradient operation.

\section{Experimental Setup}

\subsection{Datasets}
\label{sec:datasets}

\subsubsection{Training Datasets} Our model training is supported by various datasets, which can be generally classified into three categories: speech, music, and general sounds. For speech, we utilize the GigaSpeech (GGS)~\cite{chen2021gigaspeech} dataset, a comprehensive English speech recognition corpus with around $10,000$ hours of transcribed audio, and the speech dataset collected by VoiceFixer~\cite{liu2021voicefixer} on OpenSLR\footnote{\url{https://openslr.org/}}, featuring a multi-lingual speech dataset with $186,514$ short audio clips. The music category includes the Million Song Dataset (MSD)\cite{bertin2011million}, which provides a vast collection of $510,000$ music tracks with metadata. We also adopt datasets including MedleyDB~\cite{bittner2014medleydb}, and MUSDB18~\cite{musdb18} training subset, mostly used for music source separation~\cite{kong2021decoupling}.
For general sounds, we engaged with AudioSet (AS)\cite{gemmeke2017audio}, the largest classification dataset offering two million ten-second audio clips across $527$ categories, WavCaps\cite{mei2023wavcaps} that includes ChatGPT-assisted weakly-labeled audio captions for $403,050$ clips, and VGGSound (VS)\cite{chen2020vggsound}, a substantial audio-visual dataset with around $190,000$ videos from which we only utilized the audio data. All the audio data are \textcolor{black}{in $16$ bits} and resampled to $16$~kHz during training and evaluation. 
\textcolor{black}{Therefore, all SemantiCodec configurations presented in this work operate at a $16$~kHz sampling rate. We plan to explore higher sampling rate versions of SemantiCodec in future research.}
Our ablation studies in \Cref{sec-acoustic-representation-learning} and \Cref{sec-learnable-semantic-codebook} utilize \textcolor{black}{only} $10$\% of the training data to speed up the model training.

\subsubsection{Reconstruction Performance Evaluation} To assess the reconstruction capabilities of our audio codec, we carefully select and evaluate three distinct categories of datasets similar to building the training set. Within the speech category, we choose a subset of the LibriTTS clean test set~\cite{zen2019libritts}, selecting $300$ speech utterances randomly. These utterances, each lasting between $8$ and $10$ seconds, come with detailed transcription annotations, offering a rich basis for evaluating speech reconstruction accuracy.
We randomly select $500$ segments from the AudioSet evaluation set for general sound evaluation, ensuring a diverse representation of ambient sounds, effects, and non-musical content.
Our music data evaluation leverages the MUSDB18 test set~\cite{musdb18}, comprising $50$ songs with isolated tracks for vocals, drums, bass, and other elements.
From each of the four tracks and their mixture for every song, we randomly select a $10$-second segment, ensuring it is non-silent, to form our evaluation set.
Our evaluation dataset encompasses $1050$ audio samples, achieving a relatively balanced distribution across speech, music, and general sounds. 
% TODO double check
Our MUSHRA (MUltiple Stimuli with Hidden Reference and Anchor) test~\cite{mushra-series2014method} is performed on the same evaluation dataset while only $10\%$ of the data is used to save effort for subjective evaluation. Our evaluation set and metrics implementations are publicly available\footnote{\url{https://zenodo.org/records/11047204}}.

\subsubsection{Semantic Information Evaluation} To assess the richness of semantic information captured by audio codec representations, we employ a diverse array of datasets following a subset of the \textcolor{black}{Holistic Evaluation of Audio Representations}~(HEAR) benchmark~\cite{turian2022hear}.
We choose the tasks oriented towards \textcolor{black}{audio classification}~\cite{10258355}, eliminating tasks like Gunshot Triangulation, which aims to recognize recording devices.
Also, for convenience, we choose the datasets where the duration of each audio clip is shorter or close to $10$ seconds.
The evaluation datasets we employed include the following: 
(i) NSynth Pitch (NSPitch)~\cite{engel2017neural} is utilized to evaluate the ability to recognize musical pitches, a fundamental aspect of music theory and auditory perception.
(ii) ESC-50~\cite{piczak2015esc} encompasses a broad array of environmental sounds such as rainstorms and animal calls, testing the versatility of the codec features towards general sound effects.
(iii) LibriCount (LbCount)~\cite{stoeter2018libricount} is the dataset used to determine the number of speakers in an audio clip, combining speech detection with the differentiation of speakers within complex auditory scenes.
(iv) CREMA-D (CRM-D)~\cite{cao2014crema} focuses on speech emotion recognition, requiring models to classify a wide spectrum of emotional states conveyed through speech.
(v) Vocal Imitations (VoImit)~\cite{kim2018vocal} dataset examines the ability to classify non-verbal human vocal imitations of various sounds.
(vi) Speech Commands (SC)~\cite{warden2018speech} dataset tests the recognition of specific spoken commands, emphasizing speech clarity and command accuracy.
These audio classification datasets collectively offer a comprehensive evaluation framework, spanning musical notes, environmental sounds, speech nuances, and non-verbal vocalizations, to thoroughly assess the semantic capabilities of audio codec representations.

\subsection{Baselines}
\label{subsec:baselines}

We employ several state-of-the-art neural audio codecs as baselines for comparison, including Encodec~(EC), Descript Codec~(DAC), and HiFi-Codec~(HC), which have demonstrated success in the domain of general \textcolor{black}{audio}. The open-sourced Encodec\footnote{\url{https://github.com/facebookresearch/encodec}} is trained on \textcolor{black}{a variety of audio datasets sampled at $24$~kHz} and can compress audio to $1.5$, $3.0$, $6.0$, $12.0$, and $24.0$ kbps. Since SemantiCodec operates at a relatively low bitrate, \textcolor{black}{we only include Encodec at $1.5$, $3.0$ and $6.0$ kbps for comparisons}. For DAC, we adopt the open-sourced $6.0$~kbps model that operates on $16$~kHz as one of the baselines. \textcolor{black}{As DAC does not have checkpoints open-sourced for lower bitrates, we train three new versions of DAC on the same training data as SemantiCodec} following the setting described in their paper~\cite{kumar2024high}. The three DAC settings have a bit rate of $0.47$, $0.78$, and $1.41$ kbps, which are comparable to the three variants of SemantiCodec.
For HiFi-Codec, we adopt the $2.0$ kbps checkpoint\footnote{\url{https://github.com/yangdongchao/AcademiCodec}}. \textcolor{black}{The baseline models we used, including Encodec, Descript Codec, and HiFi-Codec, contain approximately $23$, $74$, and $63$ million parameters, respectively.}

\subsection{Evaluation Metrics}

\subsubsection{Reconstruction Performance Evaluation} To assess the reconstruction quality of audio codecs with objective metrics, we employ mel spectrogram distance (MEL), short-time Fourier transform distance (STFT), and the virtual speech quality objective listener score~(ViSQOL)~\cite{hines2015visqol}. MEL and STFT metrics quantify spectrogram discrepancies, with STFT providing a more nuanced capture of high-frequency fidelity. Our implementation for MEL and STFT follows the approach detailed in DAC~\cite{kumar2024high}, employing multiple window lengths to accommodate various audio signal types effectively. ViSQOL, a full-reference and intrusive perceptual quality metric, is designed to estimate the subjective mean opinion score (MOS) of audio quality. ViSQOL offers two modes for evaluation: an audio mode for $48$~kHz samples and a speech mode for $16$~kHz samples. Considering our focus on $16$~kHz audio codecs, we resample both the ground truth and reconstructed audio to $48$~kHz to leverage the audio mode for evaluations on the AudioSet and MUSDB18 datasets. Additionally, to evaluate the intelligibility of speech signals, we employ the word error rate (WER), a critical metric for evaluating the performance of automatic speech recognition systems by calculating the percentage of errors in the form of substitutions, deletions, or insertions relative to a reference transcript. We utilize the \textit{whisper-large-v3}~\cite{radford2023robust}\footnote{\url{https://huggingface.co/openai/whisper-large-v3}} model to transcribe our reconstruction and remove all punctuations in the original LibriTTS transcriptions before calculating WER.

\subsubsection{Semantic Information Evaluation} We assess the semantic distinctiveness of the audio codec representation by analyzing classification accuracy on the datasets described in Section~\ref{sec:datasets}, following the methodology outlined in the HEAR evaluation benchmark~\cite{turian2022hear}.
Our evaluation specifically focuses on the quantized output of the encoder, which serves as input to the decoder.
The codec model is frozen to extract the quantized features without fine-tuning.
The feature sequences are averaged along the time axis to obtain the clip-level features, \textcolor{black}{which represents the global information of an audio clip.}
A shallow downstream multi-layer perceptron (MLP) classifier with two linear layers is trained on clip-level features.
The classification performance is reported to indicate the semantic distinctiveness of the codec.
Furthermore, several studies on audio language models use powerful \textcolor{black}{auto-regressive}~(AR) language models to predict tokens from the first VQ layer while leaving the rest of the tokens to be predicted by non-AR language models~\cite{wang2023neural-valle}.
Therefore, we also report the classification performance using tokens from the first VQ layer to assess the semantic richness.   
% For classification purposes, we design a two-layer linear classifier that operates directly on the quantized encoder features.
% We perform average pooling on the quantized feature to form the input to the linear classifier

\subsubsection{MUSHRA Test} 
% To assess the subjective audio reconstruction quality of various audio codecs, we adhere to the established MUSHRA test protocol~\cite{mushra-series2014method}. 
\textcolor{black}{To assess the subjective audio reconstruction quality of various audio codecs, we adhere to the established MUSHRA test protocol~\cite{mushra-series2014method}. Participants are required to compare open audio samples against corresponding hidden references, predefined anchors, and the output of different systems.}
% , ensuring evaluators are unaware of the original clip to foster impartial assessments. 
Ratings are assigned on a scale from $0$ to $100$.
For our MUSHRA evaluation, we randomly select $10$\% of the audio from our evaluation set, comprising $25$ music tracks, $30$ speech recordings, and $50$ general sound samples. Participants are tasked with rating the audio quality of nine different samples, which include three variations each of SemanticCodec and Descript codecs at different bitrates, an open-source Encodec at $1.5$~kbps, the original audio (ground truth), and an anchor \textcolor{black}{of the original audio} with a low-pass filter applied at $3.5$~kHz.
\textcolor{black}{Note that the original MUSHRA test requires two anchor samples with cut-off frequency of $3.5$~kHz and $7$~kHz. However, since our test samples contain frequency information only up to $8$~kHz, the $7$~kHz anchor would not offer sufficient differentiation and was therefore omitted as an anchor in our test.}
% TODO figure draw anchor
We ensure each set of audio receives evaluations from at least $10$ raters. Participants are instructed with \textit{This task evaluates the quality proximity between an audio sample and its reference. Please listen carefully to the reference audio and then rate the quality of each test audio clip compared to the reference. Use the scale where $0$ indicates no resemblance to the reference, and $100$ means perfectly the same as the reference.}
% Ratings with ground truth scored below \textcolor{black}{$90$} are excluded to maintain data integrity. 
Including deliberately degraded anchors allows the MUSHRA test to reveal subtle perceptual distinctions and quality variances across codecs. 
\textcolor{black}{We perform post-screening following the standard MUSHRA test protocol~\cite{mushra-series2014method}.}
Our MUSHRA test has received a favourable opinion from the ethics review through completing the University of Surrey self-assessment governance and ethics form.
We calculate and report the mean MUSHRA score for each codec as its final subjective evaluation metric.

\subsection{Implementation Details}
\label{subsec:implementation_details}

% \usepackage{multirow}
% \usepackage{multirow}
% \usepackage{multirow}

We implemented the k-means clustering on the high-dimensional AudioMAE features extracted from our \textcolor{black}{training dataset mentioned} in Section~\ref{sec:datasets}. This procedure was carried out individually for three types of audio: music, speech, and general sound. For each type of audio, we run the clustering with different numbers of centroids, including $1024$, $2048$, $4096$, $8192$, and $16384$, to produce $15$ distinct sets of centroids. To manage the computational complexity of clustering high-dimensional features, we made several optimizations to the algorithm, which is available online\footnote{\url{https://github.com/haoheliu/kmeans_pytorch}}.

In building the final semantic codebook for SemantiCodec, we combined codebooks from each of the three audio domains. Given that general sound tends to encompass a broader array of sounds than speech and music, we assigned twice as many centroids. This created four distinct semantic codebooks with varying \textcolor{black}{number of} centroids: $4096$, $8192$, $16384$, and $32768$. For example, the codebook with $16384$ centroids includes $8192$ centroids for general sound and $4096$ centroids each for speech and music. This selection and combination process ensures our SemantiCodec has a comprehensive semantic codebook that accurately captures the diverse range of audio it processes. We repeat the above clustering and merging process for three different settings of stack factors \textcolor{black}{$K\in\{1,2,4\}$}. We randomly select and employ one of the four semantic codebooks in each training batch during model optimization. Therefore, our model supports variable vocabulary sizes and bitrates. \textcolor{black}{Table~\ref{tab-semanticodec-kbps-settings} summarizes the available settings of SemantiCodec with the corresponding codebook sizes, bit rates, and token rates.} The acoustic codebook utilizes a fixed-size codebook. We employ $32768$ centroid semantic codebook and $8192$ centroid acoustic codebook by default during model evaluation. Our experiment result indicates that more semantic centroids can moderately improve the reconstruction quality (see \Cref{fig:effect-of-vocab-size}). 

\begin{table}[htbp]
\caption{\textcolor{black}{Bit rate allocation table for semantic quantization layer (SEM) and acoustic quantization layer (ACO).}}
\label{tab-semanticodec-kbps-settings}
\begin{tabular}{cccccc}
\toprule
\multirow{2}{*}{\begin{tabular}[c]{@{}c@{}}Token/Sec\\      (SEM+ACO)\end{tabular}} & \multicolumn{2}{c}{Codebook Size} & \multicolumn{2}{c}{Kbps}         & \multirow{2}{*}{Overall kbps} \\
                                                                                    & SEM     & ACO                     & SEM    & ACO                     &                               \\
\midrule
\multirow{4}{*}{100}                                                                & $32768$   & \multirow{4}{*}{$8192$}& $0.75$ & \multirow{4}{*}{$0.65$} & $1.40$\\
                                                                                    & $16384$   &                         & $0.70$&                         & $1.35$                        \\
                                                                                    & $8192$    &                         & $0.65$ &                         & $1.30$\\
                                                                                    & $4096$    &                         & $0.60$&                         & $1.25$                        \\
\midrule
\multirow{4}{*}{50}                                                                 & $32768$   &                         \multirow{4}{*}{$8192$}& $0.375$ & \multirow{4}{*}{$0.325$} & $0.700$\\
                                                                                    & $16384$   &                         & $0.350$&                         & $0.675$                        \\
                                                                                    & $8192$    &                         & $0.325$ &                         & $0.650$\\
                                                                                    & $4096$    &                         & $0.300$&                         & $0.625$                        \\
\midrule
\multirow{4}{*}{25}                                                                 & $32768$   &                         \multirow{4}{*}{$8192$}& $0.1875$ & \multirow{4}{*}{$0.1625$} & $0.3500$\\
                                                                                    & $16384$   &                         & $0.1750$&                         & $0.3375$                        \\
                                                                                    & $8192$    &                         & $0.1625$ &                         & $0.3250$\\
                                                                                    & $4096$    &                         & $0.1500$&                         & $0.3125$    \\
\midrule
\end{tabular}
\end{table}

We utilize the AudioMAE model pretrained on AudioSet\footnote{\url{https://github.com/facebookresearch/AudioMAE}}, which features an embedding dimension of $768$. The resulting feature dimension is $K \times 768$ when stacking adjacent frames. Our implementation strategy for both the AudioMAE encoder and the diffusion decoder aligns with the approach described in AudioLDM 2. A notable modification in our setup is the adoption of a more compact T-UNet architecture with a base channel number of $64$ and a parameter number of $75$ million, in contrast to the larger version in AudioLDM 2 which features a base channel number of $128$ with $346$-million parameters. SemantiCodec is configured to operate at three different token rate settings, including $25$, $50$, and $100$ tokens per second. For each setting, the model undergoes training for $500,000$ steps on the designated training set with two A100-Amphere-$80$GB GPUs with a batch size of $48$. With a basic learning rate of $10^{-4}$, we incorporate a linear warm-up phase over the first $5,000$ steps. In the T-UNet architecture, the output from the SemantiCodec encoder, denoted as $\mE$, is integrated via cross-attention mechanisms. Given that $\mE$ lacks inherent positional information, we enrich it with fixed positional embeddings before input into the cross-attention layers.

SemantiCodec is trained on audio segments exactly $10.24$ seconds in length. To accommodate audio files of varying lengths, we have developed solutions involving either trimming or an overlap-add approach. For audio files shorter than $10.24$ seconds, we pad them to the required duration, compute the token encoding, and then remove any tokens corresponding to the padded region. For files longer than $10.24$ seconds, we segment the audio into $10.24$-second chunks without overlap, compute their token embeddings, and concatenate these segments. During inference, the decoder uses a window length of $10.24$ seconds with a $6.25\%$ overlap between consecutive windows. The audio content from overlapping segments is blended using a linear decay and gain before combining.

% For each of the three settings, we perform further fine-tuning on the speech dataset for $20,0000$ step, which we refer to as SemantiCodec-Speech.

% We implement the k-means clustering based on the audiomae feature extracted from our training set, as described in Section~\ref{sec:datasets}. We perform three separate times of k-means clustering on music, speech, and general sound, with centroid number setting of 1024, 2048, 4096, and 8192, resulting in 12 sets of k-means centroid. We implement various algorithm optimizations to support the large-scale clustering of the high-dimensional feature and our code is available online\footnote{\url{https://github.com/haoheliu/kmeans_pytorch}}. The final semantic codebook used in SemantiCodec is a combination of three codebook from each domain, in which we set the centroid from the general sound to be twice as many as speech because general sound contains more complex information comparing with speech and music dataset. Finally we have curated four different semantic codebook with centroid number of 4096, 8192, and 16384. For example, the 16384 codebook is a combination of 8192 centroid on general sound, 4097 centroids on speech and music.

\section{Results}
\label{sec: results}

% In this section, we first compare SemantiCodec with baseline codecs in terms of audio reconstruction quality and semantic richness.
% Then we perform comprehensive ablation studies on the impact of the design choices of SemantiCodec on the performance.

\subsection{Reconstruction Quality}

% TODO xnx reconstruction quality 里不提、不放 acc

\begin{table*}[tbp]
\scriptsize
\centering
\caption{Objective evaluation of SemantiCodecs and competing baseline codecs at various bitrates on speech, music and general sound. VIS stands for the ViSQOL metric.} % ACC denotes the classification accuracy with the audio token.}
\label{tab:reconstruction_result}
\begin{tabular}{ccc|ccc|ccc|cccc|c}
\toprule
\multicolumn{1}{l}{}            & \multicolumn{1}{l}{} &         & \multicolumn{3}{c|}{General Sound}                                                             & \multicolumn{3}{c|}{Music}                                                                     & \multicolumn{4}{c|}{Speech} &\multicolumn{1}{c}{\textbf{Average}}\\
\midrule
Model                           & Kbps &Token/Sec& MEL$\downarrow$                & STFT$\downarrow$                 & VIS$\uparrow$& MEL$\downarrow$                  & STFT$\downarrow$                 & VIS$\uparrow$& MEL$\downarrow$                  & STFT$\downarrow$                 & VIS$\uparrow$& WER$\downarrow$                      & VIS$\uparrow$ \\
\midrule
% SpeechTokenizer                             & $4.00$                            & $4.91$& $3.51$& $3.08$& $5.26$& $3.35$& $3.35$& $3.85$& $2.99$& $4.32$& $2.9$& & $3.58$&  &\\
GroudTruth                             & $-$                             & - & $0.0$ & $0.0$ & $4.99$ & $0.0$ & $0.0$ & $4.99$ & $0.0$& $0.0$ & $4.99$ & $2.09$ & $4.99$                         \\
\begin{tabular}[c]{@{}c@{}}SemantiCodec \\ w. GT AudioMAE \end{tabular}                             & $-$                             & - & $3.78$ & $3.89$ & $4.58$ & $3.79$ & $3.40$ & $4.56$ & $3.77$& $3.18$ & $4.71$ & $2.7$ & $4.61$                          \\
\midrule
\begin{tabular}[c]{@{}c@{}}SemantiCodec \\ w.o. Acoustic VQ\end{tabular}                             & $0.70$                             & $50$ & $7.29$ & $4.93$ & $2.43$ & $7.67$ & $4.44$ & $2.61$ & $8.68$& $4.58$ & $2.78$ & $55.6$ & $2.61$                          \\
\midrule
DAC                             & $6.00$                             &$600$& $2.91$& $3.03$& $4.36$& $2.83$& $2.90$& $4.54$& $2.79$& $2.92$& $4.71$& $3.0$ & $4.54$                          \\
\midrule
                                & $6.00$                             &$600$& $4.38$& $4.10$& $4.00$& $4.17$& $2.90$& $4.14$& $4.54$& $3.19$& $4.35$& $3.3$ & $4.16$                          \\
                                & $3.00$                             &$300$& $4.84$& $4.26$& $3.58$& $4.67$& $3.11$& $3.78$& $5.06$& $3.40$& $4.10$& $3.7$ & $3.82$\\

\multirow{-3}{*}{Encodec}       & $1.50$                           &$150$& $5.39$& $4.47$& $3.04$& $5.30$& $3.33$& $3.27$& $5.83$& $3.67$& $3.67$& $5.0$ & $3.33$                          \\
\midrule
HiFi-Codec                      & $2.00$                             &$200$& $4.35$& $3.61$& $3.11$& $4.37$& $3.11$& $3.42$& $3.93$& $2.99$& $4.18$& $3.6$ & $3.57$                          \\
\midrule
                                & $0.47$                          &$47$& $7.56$& $4.58$& $2.12$& $7.80$& $4.24$& $2.33$& $8.62$& $4.70$& $2.73$& $28.2$ & $2.39$\\
                                & $0.78$  &$78$& $6.73$& $4.41$& $2.47$& $6.44$& $3.88$& $2.80$& $6.76$& $4.13$& $3.19$& $11.6$ & $2.82$  \\
\multirow{-3}{*}{DAC}           & $1.41$  &$141$& $6.56$& $4.78$& $2.85$& $6.30$& $3.72$& $3.10$& $6.71$& $3.91$& $3.44$& $5.0$ & $3.13$  \\
\midrule
                                & $0.36$                          &$25$& $5.06$& $4.02$& $2.84$& $5.22$& $3.76$& $3.18$& $5.77$& $3.72$& $3.49$& $19.6$& $3.17$\\
                                & $0.70$  &$50$& $4.67$& $3.97$& $3.20$& $4.74$& $3.62$& $3.54$& $4.95$& $3.49$& $3.92$& $5.1$& $3.55$\\
\multirow{-3}{*}{SemantiCodec} & $1.40$  &$100$& $4.39$& $3.79$& $3.48$& $4.44$& $3.56$& $3.80$& $4.54$& $3.38$& $4.17$& $3.4$& $3.81$\\

\bottomrule

\end{tabular}
\vspace{-1em}
\end{table*}

\Cref{tab:reconstruction_result} shows the audio reconstruction quality of SemantiCodec and baselines described in \Cref{subsec:baselines} indicated by objective metrics. We use \textit{Token/Sec} to denote the number of tokens the audio signal is encoded into per second, which is crucial in audio language modelling as it influences the length of the audio sequence.
Given the variability in semantic codebook sizes, models with \textcolor{black}{identical token rate may yield different bit rates~(see \Cref{fig:effect-of-vocab-size}), denoted by \textit{kbps}.}
% First, for codecs operating at various bitrates (i.e., Encodec, DAC and SemantiCodec), the reduction in bitrate uniformly results in a decrease in reconstruction quality.
% As expected, fewer codebooks used in RVQ leads to a loss of finer acoustic details during quantization.
% In SemantiCodec, a different approach to reducing the bitrate is employed by increasing the stacking factor $K$.
% but shorter sequence lengths still lead to inferior reconstruction quality. 
SemantiCodec \textcolor{black}{significantly outperforms DAC on the reconstruction quality} at similar bitrates. 
\textcolor{black}{With as low bitrate as $0.70$~kbps, SemantiCodec still outperforms the $1.5$~kbps Encodec on reconstruction quality}. It is remarkably comparable to HiFi-Codec operating at $2.0$~kbps, as the average ViSQOL score indicates. 
With an ultra-low bitrate of $0.36$~kbps, SemantiCodec still achieves a better ViSQOL score than the $1.41$~kbps DAC. The $1.40$ kbps SemantiCodec demonstrates performance on par with the $3.0$ kbps Encodec.
The superior performance of SemantiCodec in reconstruction quality at low bitrates indicates its potential in efficient audio transmission, storage and audio-based language modelling since it provides shorter discrete representations of audio without substantially compromising the reconstruction quality.

Using the ground truth unquantized AudioMAE features as the condition for the latent diffusion model, the setting SemantiCodec~\textit{w. GT AudioMAE} marks the performance upper bound of SemantiCodec. SemantiCodec~\textit{w. GT AudioMAE} demonstrates strong performance on the reconstruction quality with an average ViSQOL score of $4.61$, which indicates that the original AudioMAE features contains sufficient information to reconstruct the audio signal. 
% However, the metrics MEL and STFT are still considerably lower than other neural codecs, which is expected because the generative nature of SemantiCodec weight perceptual quality more than the fine-grained details compared with other neural codecs, which mostly directly employ MEL and STFT as their training objective.
% However, the metrics MEL and STFT are still considerably lower than other neural codecs, which is expected because SemantiCodec does not directly employ MEL and STFT as the training objective.
% \textcolor{black}{Most previous works, such as DAC and Encodec, use MEL and STFT directly as training objectives, which can potentially lead to better performance on MEL and STFT. Although SemantiCodec does not use MEL and STFT as direct training objectives, it still achieves better results at comparable bitrates than DAC, which further indicates the better performance of SemantiCodec.}
\textcolor{black}{While DAC and Encodec directly use MEL and STFT as training objectives, potentially enhancing their performance on these metrics, SemantiCodec, despite not doing the same, achieves superior results at comparable bitrates, highlighting its overall better performance.}

By removing the second VQ layer, as shown in the setting of SemantiCodec \textit{w.o. Acoustic VQ}, the model performance exhibit a considerable degradation, with a WER of $55.6$ and a ViSQOL of $2.61$, highlighting the importance of the second acoustic VQ layer for acoustic reconstruction.
% However, the ACC in this case achieves a high performance of $53.1$, which indicates this layer is crucial for semantic information.
Finally, comparing the reconstruction performance across different audio types, it can be concluded that general sound consistently exhibits a slightly inferior reconstruction quality than music and speech. 
The challenges in reconstructing general sound signals may be attributed to the intrinsic complexity and variability of general sound content. This also validates our choice of assigning more semantic centroid numbers for general sound than speech and music in \Cref{subsec:implementation_details}.

% Furthermore, we modify SemantiCodec by using the unquantized, ground truth AudioMAE features and excluding the acoustic encoder and VQ layer, respectively.

\begin{figure}[tbp]
    \centering
    \includegraphics[width=1.0\linewidth]{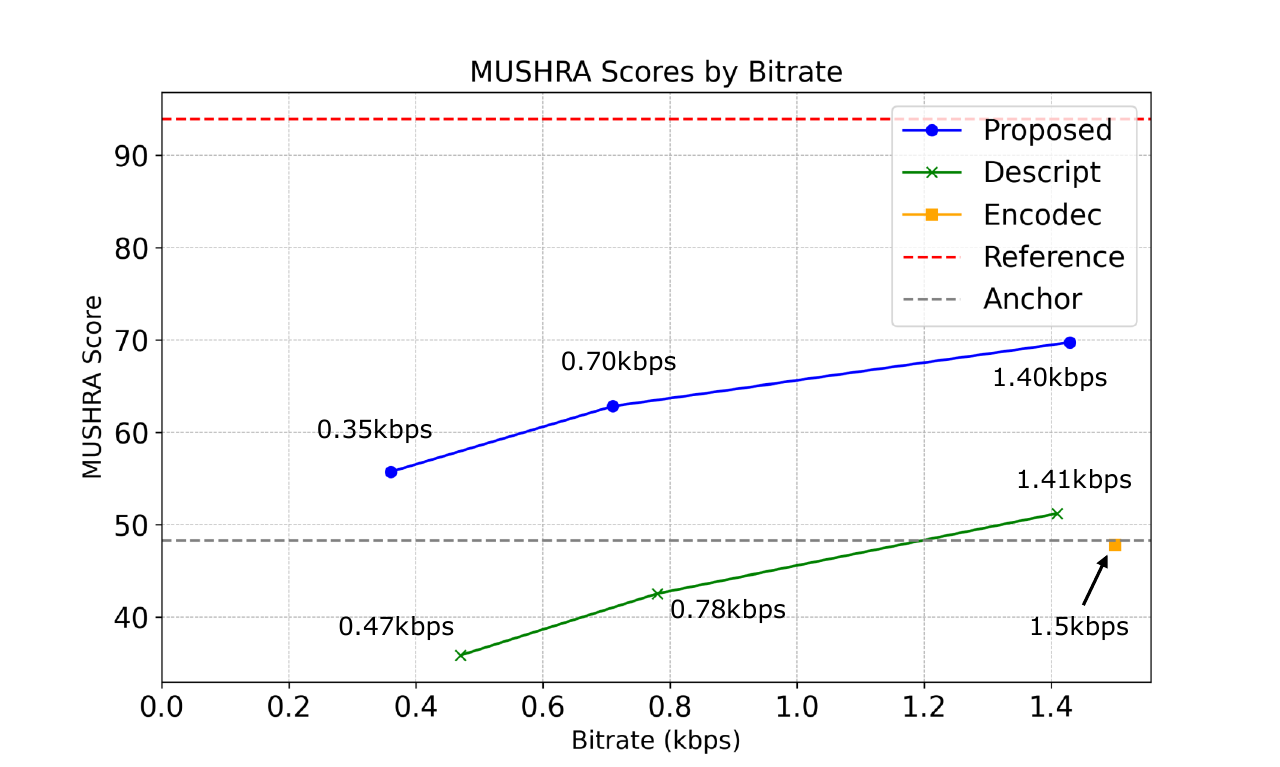}
    \caption{The average MUSHRA test score on our curated evaluation set. Our proposed SemantiCodec outperforms the baseline models even with a significantly lower bitrate. \textcolor{black}{For fair comparisons, DAC and HiFi-Codec are not included as they use much higher bit rates.}}
    \label{fig:mushra}
    \vspace{-1.5em}
\end{figure}

\begin{figure}[tbp]
    \centering
    \includegraphics[width=1.0\linewidth]{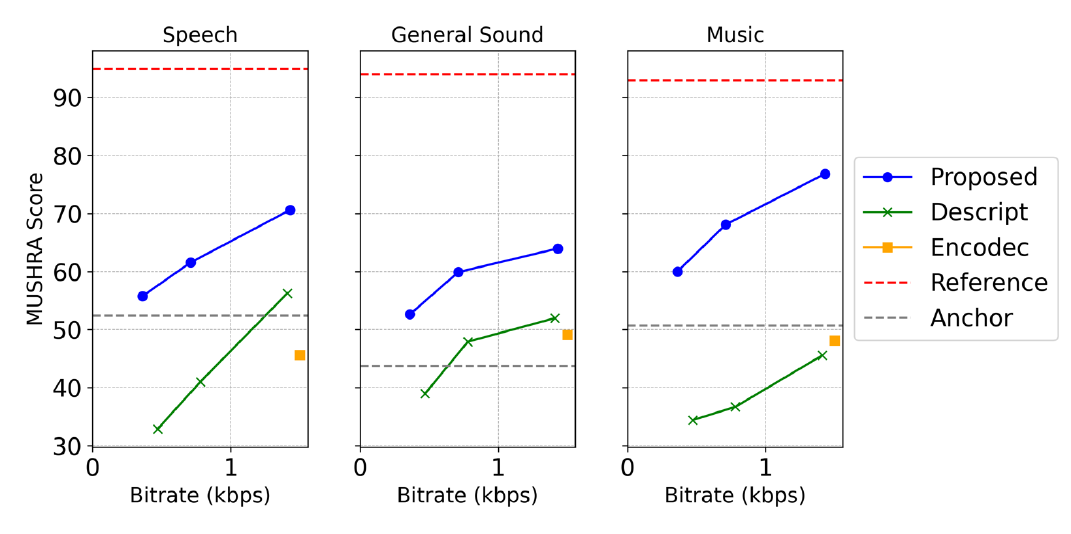}
    \caption{\textcolor{black}{The MUSHRA test score on different domains.}}
    \label{fig:mushra-audiotype}
    \vspace{-1.5em}
\end{figure}

\begin{figure*}[tbp]
    \centering
    \includegraphics[width=1.0\linewidth]{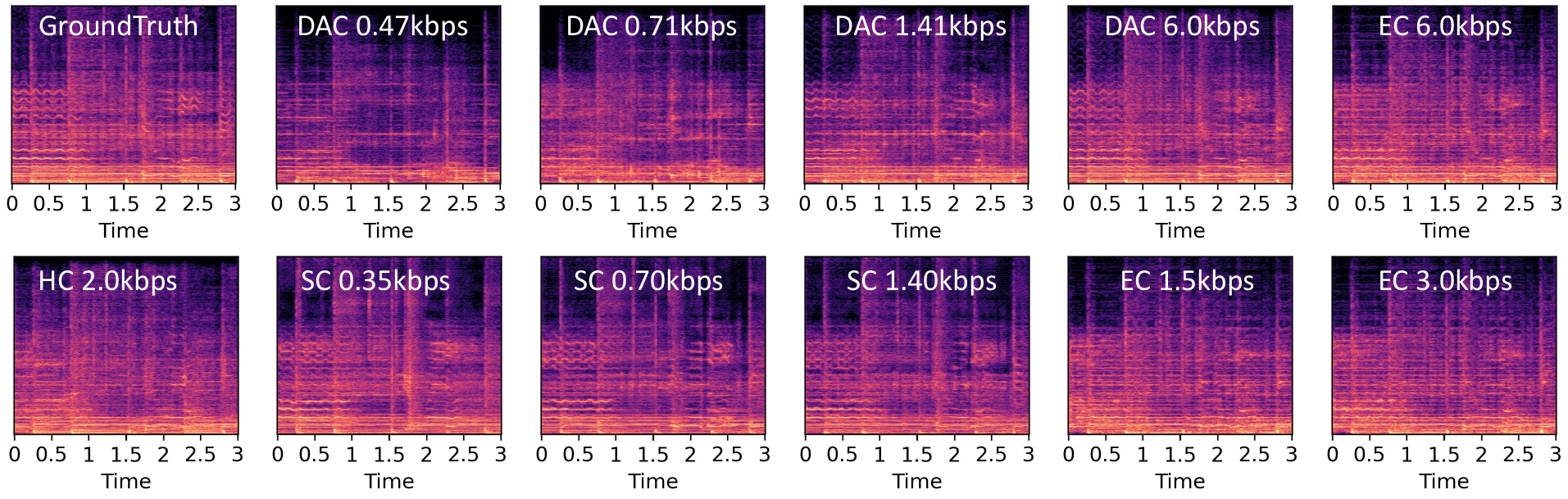}
    \caption{The log-STFT spectrogram of the ground truth audio and the reconstruction audio with different audio codecs. DAC, EC, HC, and SC are the descript codec, Encodec, HiFi-Codec, and SemantiCodec, respectively.}
    \label{fig:demo}
\end{figure*}

\Cref{fig:mushra} showcases the reconstruction performance in terms of subjective MUSHRA scores.
We only incorporate codecs with a bitrate of less than or equal to $1.5$~kbps into evaluation since we aim at developing codecs with ultra-low bitrates.
The result is consistent with the objective evaluation shown in \Cref{tab:reconstruction_result} that SemantiCodec significantly outperforms counterparts with a similar or even higher bitrate. With the reference audio being scored to $93.9$, the $1.5$~kbps Encodec model achieves a $47.8$ average MUSHRA score, while even our $0.35$ kbps SemantiCodec achieves a significantly higher score of $55.7$. The high MUSHRA score of SemantiCodec can be attributed to the generative nature of the model, which leads to better audio quality at ultra-low bitrate.
By comparison, we observe strong artifacts at similar bitrates on DAC and Encodec (shown in \Cref{fig:demo}), which can significantly impact \textcolor{black}{the perceived quality for the listener}. 
\textcolor{black}{\Cref{fig:mushra-audiotype} provides further analysis of the MUSHRA score on different domains of audio, which shows similar trends as \Cref{fig:mushra}. SemantiCodec demonstrates significantly better performance in music, typically involving more complex acoustic information. This suggests that SemantiCodec is highly effective at handling intricate scenarios.}

\subsection{Semantic in the Codec Tokens}
\label{subsec:semantic_result}

\begin{table*}[htpb]
    \centering
    \scriptsize
    \caption{Semantic evaluation of SemantiCodecs and competing baseline codecs using the HEAR benchmark.}
  \begin{tabular}{c|ccc|cccccc|c}
                \toprule
        VQ setting & Model & Kbps & Token/Sec & ESC-50~(\%) & NSPitch~(\%) & SPC~(\%) & LbCount~(\%) & CRM-D~(\%) & VoImit~(\%) & Average~(\%)\\
        \midrule
        Unquantized & AudioMAE & -& -& $79.5$& $82.0$& $48.0$& $69.4$& $67.3$& $17.4$& $60.6$\\
         \midrule
    \multirow{9}{*}{All VQ layers} & Encodec & $6.00$& $600$& $40.7$& $60.8$& $27.3$& $45.0$& $44.2$& $4.4$& $37.1$\\
     & HiFi-Codec & $2.00$& $200$& $36.3$& $71.0$& $26.5$& $58.2$& $45.6$& $4.3$& $40.3$\\
     \cline{2-11}
     & \multirow{4}{*}{DAC}& $6.00$& $600$& $33.4$& $56.1$& $21.0$& $46.4$& $39.9$& $3.4$&$33.4$\\
     & & $1.41$& $141$& $41.1$& $\mathbf{80.9}$& $30.3$& $59.5$& $44.7$& $4.8$& $43.5$\\
     & & $0.78$& $78$& $39.5$& $78.5$& $30.3$& $59.5$& $45.7$& $5.0$& $43.0$\\
     & & $0.47$& $47$& $36.7$& $75.7$& $26.4$& $59.6$& $44.5$& $4.9$& $41.3$\\
     \cline{2-11}
     & \multirow{3}{*}{SemantiCodec} & $1.40$& $100$& $\mathbf{63.8}$& $73.3$& $\mathbf{43.6}$& $67.0$& $\mathbf{57.9}$& $\mathbf{9.6}$& $\mathbf{52.5}$\\
     & & $0.70$& $50$& $60.9$& $64.9$& $41.9$& $\mathbf{71.6}$& $53.2$& $9.3$&$50.3$\\
     & & $0.35$& $25$& $56.4$& $61.3$& $33.7$& $70.4$& $46.9$& $8.0$&$46.1$\\
 \midrule
    \multirow{9}{*}{First VQ Layer} & Encodec & $0.75$& $75$& $32.0$& $45.3$& $23.0$& $44.8$& $40.7$& $4.2$& $31.6$\\
     & HiFi-Codec & $1.00$& $100$& $33.7$& $58.9$& $25.9$& $58.3$& $44.3$& $4.1$& $37.5$\\
     \cline{2-11}
     & DAC ($6.00$k) & $0.5$& $50$& $23.8$& $23.3$& $15.9$& $43.1$& $38.4$& $3.1$&$24.6$\\
     & DAC ($1.41$k) & $0.16$& $16$& $29.0$ & $44.2$ & $19.9$ & $57.0$ & $39.5$ & $4.0$ & $32.3$ \\
     & DAC ($0.78$k) & $0.16$& $16$& $27.6$ & $39.9$ & $17.9$ & $56.6$ & $40.9$ & $4.0$ & $31.1$ \\
     & DAC ($0.36$k) & $0.16$& $16$& $29.7$ & $46.9$ & $18.3$ & $58.2$ & $43.0$ & $4.0$ & $33.3$ \\
     \cline{2-11}
     & \multirow{3}{*}{SemantiCodec} & $0.70$& $50$& $\mathbf{66.6}$& $\mathbf{73.9}$& $\mathbf{42.7}$& $\mathbf{66.7}$& $\mathbf{57.5}$& $\mathbf{11.1}$& $\mathbf{53.1}$\\
     & & $0.36$& $25$& $64.4$& $70.2$ & $36.0$& $65.0$& $54.1$& $10.7$&$50.1$\\
     & & $0.18$& $13$& $59.6$& $66.3$& $30.7$ & $61.3$& $45.8$& $9.8$&$45.6$\\
  \bottomrule
    \end{tabular}
    \label{tab:semantic_result}
\end{table*}

The semantic richness of different codecs is demonstrated in \Cref{tab:semantic_result}.
First, when all VQ layers are utilized, SemantiCodec significantly outperforms baseline models in semantic information.
Notably, even at a low bitrate of $0.35$~kbps, the semantic performance of SemantiCodec surpasses that of higher bitrate counterparts like the $6.0$~kbps Encodec and DAC.
This indicates that AudioMAE features serve as effective means of semantic encoding.

Then, comparing codecs at different bitrates below $1.5$~kbps, both DAC and SemantiCodec exhibit a decline in semantic performance as the bitrate decreases.
However, even at the lowest bitrate of $0.35$~kbps, SemantiCodec outperforms DAC at $1.41$~kbps.
Interestingly, despite the inferior reconstruction quality of DAC at lower bitrates, the semantic richness retained at low bitrates is significantly better than that of $6.0$~kbps.
This may be attributed to the information bottleneck imposed on the encoded representations during training with low bitrates.
The model must preserve coarse audio contents in representations while discarding acoustic details, making the classifier easier to train on downstream tasks.

Since many studies utilize tokens from the first VQ layer to train audio language models, we investigate the semantic information in tokens under this situation.
As shown in the lower part of~\Cref{tab:semantic_result}, the baseline codecs exhibit a substantial drop in semantic performance when switching from \textcolor{black}{evaluating all VQ layers to the first layer only}. For example, the first VQ layer of the $6.0$~kbps DAC only achieves an average accuracy of $24.6$.
In contrast, SemantiCodec maintains similar semantic accuracy with only the first VQ layer.
This supports our assumption that semantic information is primarily encoded by semantic codes obtained through k-means clustering of AudioMAE features, with acoustic details augmented by subsequent acoustic codes.
% Using only the first VQ layer, SemantiCodec not only vastly outperforms baselines in terms of semantic richness, but also reduces the frame rate to $50$.
However, there is still a performance gap between unquantized AudioMAE features~(ACC $60.6$) and the quantized latent space of SemantiCodec~(ACC $53.1$), highlighting the need for further research into mitigating the loss caused by quantization.

% Param counts
% Encodec 6kbps

\subsection{Ablation Study}
\begin{figure*}[tbp]
    \centering
    \includegraphics[width=1.0\linewidth]{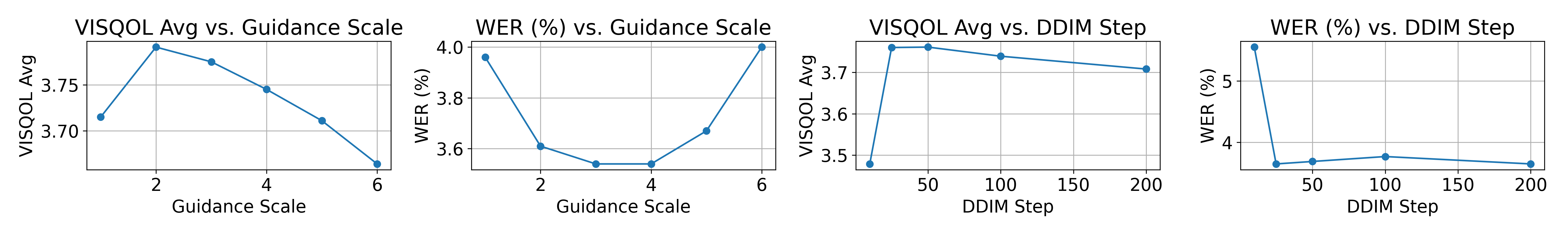}
    \caption{The impact of CFG guidance scale and DDIM sampling steps on reconstruction quality.}
    \label{fig:ablation}
\end{figure*}

\begin{table}[tbp]
\centering
\caption{Utilizing a variable semantic codebook size is beneficial for the reconstruction quality, compared with using a fixed semantic codebook \textcolor{black}{of} size $32768$.}
\label{tab:reconstruction-quality-variable-semantic-codebook}
\begin{tabular}{ccccc}
\toprule
\multicolumn{1}{l}{}        & \multicolumn{2}{c}{fixed vocab size} & \multicolumn{2}{c}{variable vocab size} \\
\midrule
Stack Factor                & ViSQOL-Avg                  & WER                   & ViSQOL-Avg                 & WER                 \\
\midrule
$K=1$                         & $3.78$                    & $3.60$& $3.81$                   & $3.40$\\
$K=2$ & $3.41$                    & $6.30$& $3.55$                   & $5.10$\\
$K=4$                         & $3.04$                    & $21.8$                  & $3.17$                   & $19.6$ \\
\midrule
\end{tabular}
\end{table}

\subsubsection{Variable Semantic Codebook Size} As detailed in Section~\ref{subsec:implementation_details}, during training, we employ a variety of semantic codebooks, enabling SemantiCodec to accommodate different vocabulary sizes. To explore the efficacy of this approach, we conduct experiments comparing a fixed vocabulary size against a variable one, maintaining \textcolor{black}{an acoustic codebook size of $8192$} in both scenarios. We \textcolor{black}{also} investigate different stack factors \(K\) to assess their impact on performance. As presented in Table~\ref{tab:reconstruction-quality-variable-semantic-codebook}, utilizing a variable vocabulary size significantly enhances the average ViSQOL score and reduces the WER. This improvement likely stems from the exposure to diverse codebooks during training, which enhances the ability of the model to interpret and process the quantized AudioMAE features effectively.

\begin{table}[t]
    \centering
    \caption{The impact of k-means centroid number on the semantic richness of SemantiCodec.}
    \begin{tabular}{ccc}
    \toprule
    K-means Centroids   & ACC \\
    \midrule
    $4096$ & $49.0$ \\
    $8192$ & $50.9$ \\
    $16384$  & $52.1$ \\
    $32768$ & $\mathbf{53.1}$ \\
    \bottomrule
    \end{tabular}
    \label{tab:kmeans_semantic}
\end{table}

\begin{figure}[tbp]
    \centering
    \includegraphics[width=1.0\linewidth]{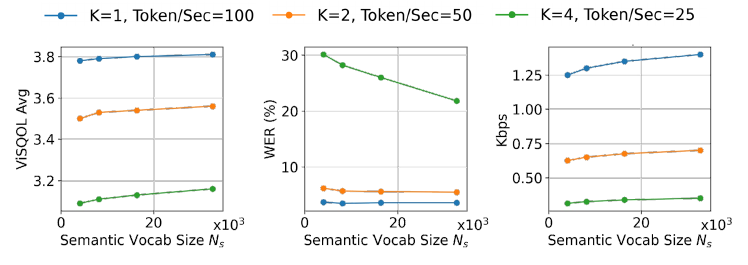}
    \caption{The impact of semantic vocabulary size on average ViSQOL, WER, and bitrate. }
    \label{fig:effect-of-vocab-size}
\end{figure}

We also study the effect of semantic codebook size on semantic richness by calculating classification accuracy with the quantized AudioMAE features. \Cref{tab:kmeans_semantic} shows as the centroid number increases, the quantization error induced by k-means modelling is reduced, hence the semantic performance consistently improves. 
Similar behaviour can be observed in the reconstruction performance, as shown in \Cref{fig:effect-of-vocab-size}. 
\textcolor{black}{When using a larger number of k-means centroids} (i.e., semantic vocabulary size), reconstructed speech quality steadily improves. However, as \textcolor{black}{also} shown by \Cref{fig:effect-of-vocab-size}, higher semantic vocab size \textcolor{black}{will} impact the bitrate, posing a trade-off between reconstruction quality and bitrate.

% As previously illustrated, the semantic richness of SemantiCodec is primarily determined by the first VQ layer.
% Therefore, in terms of semantic evaluation, we only investigate the impact of the k-means clustering settings while for reconstruction quality evaluation, we analyze the effect of several design choices and inference settings.


% \subsubsection{Encoder design}

\subsubsection{Acoustic Representation Learning}
\label{sec-acoustic-representation-learning}
The size of the acoustic codebook determines the granularity of the trainable vector quantization module.
Analogous to the influence of k-means centroid numbers, larger codebook size leads to better reconstruction quality, as indicated by \Cref{tab:encoder_reconstruct_ablation}.

The complexity of the LSTM acoustic encoder influences its capacity to refine and augment quantized AudioMAE features.
Replacing the original bi-directional LSTM with a unidirectional one results in a significant drop in speech reconstruction quality, as indicated by WER.
This validates the necessity of the bidirectional connection in the LSTM for efficiently extracting and encoding the contextual information and dependencies of AudioMAE feature sequences.

\begin{table}[tbp]
\small
\centering
\caption{The impact of SemantiCodec encoder parameters on the reconstruction quality.}
\label{tab:encoder_reconstruct_ablation}
\begin{tabular}{cccc}
\toprule
 \begin{tabular}[c]{@{}c@{}} Acoustic Codebook 
Size\end{tabular}& \begin{tabular}[c]{@{}c@{}}LSTM \\ Bi-directional\end{tabular}& \begin{tabular}[c]{@{}c@{}}VISQOL\\ Avg\end{tabular} & \begin{tabular}[c]{@{}c@{}}WER \\ (\%)\end{tabular}\\
\midrule
% $4096$             & $8192$          & \checkmark                   & $3.572$      & $4.84$     \\
% $8192$             & $8192$          & \checkmark                   & $\mathbf{3.608}$      & $4.62$     \\
% $16384$             & $8192$          & \checkmark                   & $3.605$      & $4.59$     \\
 $8192$          & \checkmark                   & $\mathbf{3.602}$      & $\mathbf{4.57}$     \\
 $4096$          & \checkmark                   & $3.583$      & $5.05$     \\
 $2048$          & \checkmark                   & $3.550$      & $4.66$     \\
 $8192$          & \xmark                   & $3.596$      & $4.80$    \\
\bottomrule
\end{tabular}
\end{table}

\subsubsection{Learnable Semantic Codebook}
\label{sec-learnable-semantic-codebook}
\begin{table}[tbp]
\centering
\caption{The impact of replacing the k-means centroid with a learnable vector quantization layer.}
\label{tab-semanticVQ}
\begin{tabular}{cccc}
\toprule
SemantiCodec          & \multicolumn{1}{c}{ViSQOL}$\uparrow$ & \multicolumn{1}{c}{WER}$\downarrow$ & \multicolumn{1}{c}{ACC}$\uparrow$ \\
\midrule
Original              & $\mathbf{3.60}$                        & $\mathbf{4.57}$                    & $\mathbf{53.1}$                    \\
Learnable Semantic VQ & $3.57$                       & $5.10$                     & $28.5$       \\        
\bottomrule
\end{tabular}
\end{table}

The semantic codebook in SemantiCodec is frozen during model training.
To validate the importance of performing k-means clustering beforehand, we conduct another experiment by replacing the semantic codebook with a learnable VQ layer, similar to the acoustic VQ layer.
As shown in~\Cref{tab-semanticVQ}, even though the reconstruction performance of SemantiCodec is not largely affected without the use of k-means centroids, the semantic evaluation accuracy shows a notable decrease from $53.1$\% to $28.5$\%.
% This indicate the k-means centroids used in the semantic codebook is crucial to maintain good semantic structure in the audio tokens.
This indicates the validity of k-means centroids used in the semantic codebook for maintaining rich semantic information in audio tokens.
Directly training a codec model with two VQ layers results in the codec neglecting semantic information and focusing solely on reconstruction capabilities, as the training loss function is limited to reconstruction loss.
% Therefore, the impact on reconstruction performance is slight but there is a significant decline in semantic richness.
% Conversely, employing frozen k-means tokens effectively preserves the semantic information learned within AudioMAE.
% The SemantiCodec encoder and decoder are utilized to augment the model's reconstruction abilities.

\subsubsection{DDIM Sampling Setups}
As shown in \Cref{fig:ablation}, the guidance scale $w$ (see \Cref{eq:cfg}) in CFG can influence the reconstruction quality.
CFG does not provide enough condition-oriented guidance through unconditional generation probability when the scale is too small.
Conversely, when the scale is too large, the effect of the condition is diluted by the emphasis towards unconditional generation probability.
Therefore, $w$ with values that are too small or too large leads to unsatisfactory reconstruction quality. 
The best reconstruction quality is achieved under a moderate scale near $3.0$.

The sampling step also plays a crucial role in the reconstruction quality of our latent diffusion decoder.
The reconstruction quality of SemantiCodec is moderate, with a small number of sampling steps, e.g., ten steps. \textcolor{black}{With a number of sampling steps larger than} $25$, the reconstruction quality significantly improves. 
% With $25$ sampling steps and a classifier-free guidance scale of $2.0$, SemantiCodec can be 
% Satisfactory results can be obtained with relatively few steps, and further increasing the step number does not significantly enhance the results.
This is promising for rapid sampling during audio reconstruction with tokens from SemantiCodec.

\section{Conclusion}

In this study, we have proposed SemantiCodec, an audio codec which can be applied to diverse audio types with ultra-low bitrates and rich semantic information in the audio tokens. SemantiCodec supports several bitrates between $0.31$~kbps and $1.40$~kbps. With a semantic and acoustic decoupled architecture, SemantiCodec achieves effective compression without significantly sacrificing quality with token rates of $50$ and $100$ per second. At an ultra-low rate of $25$ tokens per second, SemantiCodec still demonstrates significantly strong audio quality compared with state-of-the-art audio codecs. Our experiment result also shows that the semantic information within the SemantiCodec tokens is significantly richer than in previous neural audio codecs.
% A semantic and acoustic decoupled architecture is employed for SemantiCodec, which consists of a semantic encoder that utilizes discrete AudioMAE representations via k-means clustering and an acoustic encoder designed to capture the nuanced acoustic details that are essential for high-quality audio reconstruction.
% An LDM decoder achieves high-quality reconstruction of compressed audio, maintaining fidelity to the original recordings.
% Three variants of SemantiCodec with bitrates of $0.35$~kbps, $0.70$~kbps, and $1.40$~kbps, are presented to provide a tradeoff between compression efficiency and reconstruction quality.
% Experiments reveal that SemantiCodec shows significantly better the reconstruction quality and semantic richness over SOTA codecs, even with a much lower bitrate.
% Specifically, SemantiCodec presents better semantic richness than all codecs, even at lower bitrates.
% Our experiment shows even the SemantiCodec with $0.35$ kbps can have much better semantic structure than $6.0$ kbps SOTA codecs in the audio tokens.
% Future work will be utilizing SemantiCodec to perform efficient audio language modelling.

\section*{Acknowledgments}

This research was partly supported by the British Broadcasting Corporation Research and Development~(BBC R\&D), Engineering and Physical Sciences Research Council (EPSRC) Grant EP/T019751/1 ``AI for Sound'', and a PhD scholarship from the Centre for Vision, Speech and Signal Processing (CVSSP), Faculty of Engineering and Physical Science (FEPS), University of Surrey. This publication is supported by multiple datasets that are openly available at locations referenced in this paper. For the purpose of open access, the authors have applied a Creative Commons Attribution (CC BY) license to any Author Accepted Manuscript version arising. \textcolor{black}{We would also like to express our gratitude to the anonymous reviewers for their valuable comments.}

\bibliographystyle{IEEEtran}
\bibliography{refs}

\end{document}